\documentclass[11pt,reqno]{amsart}

\usepackage[margin=2.55cm]{geometry}
\usepackage{graphicx}
\usepackage{subcaption}

\usepackage{amsmath, amsfonts, amssymb, amsthm}
\usepackage{mathtools}
\usepackage{bbm}
\usepackage{booktabs}
\usepackage{bm}

\theoremstyle{plain}

\theoremstyle{definition}

\theoremstyle{remark}

\numberwithin{equation}{section}

\usepackage[colorlinks=true, linkcolor=blue, citecolor=red]{hyperref}
\usepackage{cleveref}
\usepackage{orcidlink}

\usepackage[draft,textwidth=2.75cm]{todonotes}
\renewcommand{\texttt}[1]{%
  \begingroup
  \ttfamily
  \begingroup\lccode`~=`/\lowercase{\endgroup\def~}{/\discretionary{}{}{}}%
  \begingroup\lccode`~=`[\lowercase{\endgroup\def~}{[\discretionary{}{}{}}%
  \begingroup\lccode`~=`.\lowercase{\endgroup\def~}{.\discretionary{}{}{}}%
  \catcode`/=\active\catcode`[=\active\catcode`.=\active
  \scantokens{#1\noexpand}%
  \endgroup
}

\begin{document}

\title{Quantifying Gerrymandering in Ohio: A Cycle Walk-based Ensemble Analysis}

\author[Herschlag]{Gregory J.\ Herschlag\textsuperscript{1}\,
\orcidlink{0000-0001-5443-6449}}
\email{gjh@math.duke.edu}

\author[Mattingly]{Jonathan C.\ Mattingly\textsuperscript{1,2}\,
\orcidlink{0000-0002-1819-729X}}

\author[Zhou]{Franklin Xiaozhou Zhou\textsuperscript{2}\,
\orcidlink{0000-0003-0618-232X}}

\address{\textsuperscript{1}Department of Mathematics, Duke University,
Durham, NC 27708-0320, U.S.A.}

\address{\textsuperscript{2}Department of Statistical Science, Duke University,
Durham, NC 27708-0251, U.S.A.}

\begin{abstract}
    Ohio adopted three congressional redistricting plans after the 2020 U.S. Census. The Supreme Court of Ohio ruled against the 2021 and 2022 plans. The U.S. Supreme Court later vacated the judgment concerning the 2022 plan and remanded the case, which the Ohio court then dismissed. The 2022 plan governed the 2022 and 2024 elections. In October 2025, the Ohio Redistricting Commission unanimously adopted the third plan.

    We evaluate the extent of partisan gerrymandering by comparing the three plans with ensembles that reflect the nonpartisan criteria observed in the official plans. These ensembles provide a baseline that accounts for Ohio's political geography, including its county and precinct structure, population density, and voting patterns. Ohio's stringent redistricting criteria are thought to be difficult to sample the space of plans efficiently. In our tests, multiscale Forest Recombination mixed poorly. However, the Cycle Walk algorithm showed good agreement across independent chains and allowed us to tune the underlying probability measure to reflect different nonpartisan redistricting policies.

    We find unusually Republican-favoring electoral outcomes in all three official plans relative to our ensembles. For the 2021 and 2022 plans, these findings are consistent with the Ohio Supreme Court's original rulings. We also examine 104 publicly submitted plans. We analyze these maps and find that, despite some being extremely gerrymandered, the median public submission plan is reasonably inline with our ensembles. 
\end{abstract}

\maketitle


\vspace{0.5cm}
Ohio's congressional map has been in legal dispute since the 2020 Census.
In 2018, Ohio voters amended the state constitution to add Article XIX.
When the General Assembly adopts a plan without bipartisan support, Article XIX forbids a plan that ``unduly favors or disfavors a political party or its incumbents'' or that ``unduly split[s] governmental units'' \cite{oh_const_art_xix_sec1}.
In January 2022, the Supreme Court of Ohio held that the General Assembly's 2021 plan violated both provisions, unduly favoring the Republican Party and splitting Hamilton, Cuyahoga, and Summit Counties \cite{adams_v_dewine_2022}.
In July 2022, it held that the replacement plan adopted by the Ohio Redistricting Commission again unduly favored the Republican Party \cite{neiman_v_larose_2022}.
Because the 2022 primary had already been held and the remedial process stalled, that plan nevertheless governed the 2022 election.
Following the U.S. Supreme Court's remand in June 2023 and the subsequent dismissal of the legal challenge in September 2023, the replacement plan also governed the 2024 election.
In October 2025, the commission adopted a third plan, intended to remain in effect through the end of the decade.

As we find in this study, the 2025 plan appears to be designed to elect ten Republicans and five Democrats.
In the 2026 election, a strong Democratic swing instead produced a delegation of seven Republicans and eight Democrats.
That outcome invites two opposite readings: that the map is fair enough to respond to voters, or that it is still a Republican gerrymander that a large swing happened to overcome.
We find evidence for the second reading.

To make this argument, we compare each of the three post-2020 plans with an ensemble of alternative plans drawn from an explicit probability measure.
The measure acts as a nonpartisan redistricting policy: it assigns each plan a relative weight, so that we can say, for example, that one plan is ten times as favorable as another under that policy.
It is built to reflect the nonpartisan criteria visible in the legislature's own maps.
Because the legislature has never stated those criteria, we infer them from the three released maps, and we require the sampled plans to match them in compactness, municipal splitting, and county preservation.
We do not track minority opportunity districts explicitly because \emph{Louisiana v. Callais} has substantially weakened the reading of Section 2 of the Voting Rights Act.
The benchmark logic is standard: if an enacted plan resembles the ensemble on nonpartisan criteria but is extreme in its partisan consequences, then geography and legal constraints do not explain those consequences \cite{katz_theoretical_2020,chin_signature_2019}.

Earlier ensemble studies of Ohio do not provide this benchmark.
Tapp et al. use Ohio's congressional districts to test how sensitive ReCom ensembles are to parameter choices, rather than to match the legislature's policy \cite{tapp_parameter_2025}, and since the ReCom Distribution is unknown, it is harder to generalize to studies with known distributions directly based on the legislature's policy choices.
Wise and Rossiter study how city and county preservation affect neutral map drawing in Ohio, using random maps that are not drawn from a known distribution \cite{wise_nosplit_2026}.
Both find that outcomes change little as policy parameters vary \cite{wise_nosplit_2026,tapp_parameter_2025}.
The ALARM project provides an ensemble for every state, including Ohio \cite{mccartan_simulated_2022}, and a follow-up study used those ensembles to measure the nationwide effect of gerrymandering \cite{kenny_widespread_2023}.
But ALARM generated its Ohio ensemble before the first post-2020 map was published, so it could not be tuned to the legislature's expressed non-partisan policies.
Its sequential Monte Carlo sampler also reduced municipal and county splits only through edge weights on the precinct dual graph.
We find that this is not enough to keep significant portions of certain municipalities together, as Section~2(B)(4) of the Ohio Constitution requires. Thus, although the partisan properties of an ensemble may not be sensitive to the specifics of the Ohio legislature's criteria, our study is able to directly test this.

The constitution that makes this comparison valuable also makes the sampling difficult.
Beyond the traditional criteria, Article XIX, Section~2(B) sets out several strict nonpartisan rules \cite{oh_const_art_xix_sec1,oh_const_art_xix_sec2}:
\begin{enumerate}
\item[\S(4)] If a municipal corporation or township exceeds the ideal district population, mapmakers must attempt to include a significant portion of it in a single district. Otherwise, a municipality or township with at least 100,000 residents and no more than the ideal district population must not be split; if several qualify, only the most populous receives that protection.
\item[\S(5)] Sixty-five counties must remain wholly within a district; eighteen may be divided between at most two districts, and five between at most three. These are allowances, not required splits.
\item[\S(6)] A district's portion within a split county must be contiguous within that county.
\item[\S(7)] Two districts cannot share portions of more than one county, except for counties exceeding 400,000 residents.
\item[\S(8)] Mapmakers must attempt to include at least one whole county in each district, except for districts contained entirely within one county or where federal law prevents it.
\end{enumerate}
Although XIX.2(B)(5) allows for some counties to be divided into three, the recent supreme court ruling declared that when specific counties were split into three, the commission unduly split political regions and the subsequent redistricting plans used in 2022 and 2026 split no county more than twice \cite{adams_v_dewine_2022}.

Wise and Rossiter note that ``Ohio's rule sets constrain both municipal and county splitting, beyond the capabilities of most simulation algorithms'' \cite{wise_nosplit_2026}.
Imposing these rules as hard constraints may cause chains to mix poorly, which makes Ohio a demanding test for any sampler.

In this paper, we explore using the Cycle Walk algorithm \cite{deford_cyclewalk_2025,cyclewalk_software,cyclewalk_tutorial}, a recent algorithm that efficiently samples a wider class of measures than ReCom \cite{deford_recombination_2021} and its Metropolized and multiscale variants \cite{autry_metropolized_2023,autry_metropolized_2021}.
We adapt Cycle Walk with bespoke constraints and subsampling to draw plans under the policies we study.
Preferences such as county and municipal preservation enter the spanning-forest measure through edge weights and other explicit modifications including modifications to energy functions on the measure.
The sampler can then favor plans that respect nested political geography and still move freely when strict preservation conflicts with other requirements.
Ohio's stringent rules make it a stress test for this approach, and we show that Cycle Walk samples the space effectively.

We use the adopt three measures, each tuned to match slightly different criteria, for our comparisons.
We build a primary ensemble which is matched to the non-partisan properties of the 2025 plan and supplement it with one tuned to the non-partisan properties of the 2021/2022 plans along with another that emphasizes keeping political regions more intact. We compare the nonpartisan and partisan properties of the three legislative plans against the three measures and resulting ensembles.
We find that the three plans put forward by the Ohio legislature were significantly less proportional and symmetric than they otherwise would have been in a neutral process.
We also show that the commission could have greatly reduced splitting municipal corporations while still maintaining other non-partisan criteria and achieving less partisan favor.
We also compare the three plans against the 104 plans that members of the public submitted to the commission, a community-drawn benchmark that few other states offer -- we find that although many of the community plans do not match the levels of compactness or preservation of municipal corporations and townships, many of the submissions appear to have been made in good faith and others appear to have been put forward with partisan intent.

\section{Determining a Redistricting Policy}
The 2020 census led to a congressional reapportionment that allotted 15 congressional seats to Ohio. When drawing the 15 corresponding districts, the Ohio constitution lists several strict criteria along with several factors that \emph{may} be accounted for. We begin by describing how the constitutional statutes may be interpreted based on Ohio's demographics and geography along with our relevant data sources. Redistricting commissions have certain latitude in how they interpret and implement certain redistricting stipulations, and we then turn to probing which non-partisan criteria the redistricting commission used. Ultimately, we develop several potential policies along with corresponding ensembles of plans that adhere to these policies.

\subsection{Ohio Redistricting Policy}
As listed above, the Ohio Constitution has an extensive list of county, municipal corporation, and township preservation criteria. Ohio law generally classifies a municipal corporation as a city if it had at least 5,000 residents at the last federal census or meets the threshold of 5,000 registered resident electors or voters under Ohio Revised Code Section~703.011. Other municipal corporations are classified as villages \cite{oh_code_municipal_classification}. The U.S. Census records municipal corporations as a subset of its Place data.

Ohio townships are unincorporated units governed by 3 elected Trustees and 1 elected Fiscal Officer. Townships are county subdivisions that cannot levy taxes and cannot pass local laws, but can self-regulate in other ways, such as regulating road or public land infrastructure \cite{oh_code_township}. Townships are collected and disseminated by the US Census as part of the County Subdivision data.
Contrasted to townships, municipal corporations do not need to be county sub-divisions, can overlap townships, and are able to annex partial or full townships. When a municipal corporation annexes a full township, the township becomes incorporated and falls under the municipal corporation's governance. This data is kept and described by the US Census which provides a State Guide for Ohio \cite{oh_geo_data}.

When considering municipal corporations and townships in the context of redistricting, there are several stringent rules (as outlined above) and other rules that are left to the interpretation of the redistricting commission. For an example of the latter, the redistricting body must ``attempt to include a significant portion of it in a single district." The Ohio Constitution provides careful definitions of municipal corporations and cities for the purpose of redistricting. For example, if a city spans two counties, the two parts of the city in separate counties are considered as two separate cities. The constitution also carefully defines how to determine whether a county, municipal corporation, or township is split, which is if ``any contiguous portion of its territory is not contained entirely within one district." However, other than the explicit criteria for preserving units larger than 100,000 people, there is no explicit constitutional mandate to preserve smaller townships or municipal corporations.

Ohio also has strict county-splitting criteria. As discussed in the introduction, the recent Ohio Supreme Court case lead, in effect, to a mandate to never split a county more than twice. This is more stringent than the constitutional code, but we adopt it in the present work.

In addition to laws preserving political boundaries, Ohio also requires congressional districts to be compact and contiguous \cite{oh_const_art_xix_sec2}. The commission does not report directly on what standards of compactness should be used, however, much of the language around compactness in \emph{Adams v. DeWine} involved using the standard measures of Polsby-Popper, Reock, and the Convex Hull Area Ratio \cite{cooper2022compactness,league2022opinion,rodden2023comparison,freilich2021testimony,ohio2021transcript}. We adopt the first two in the current work, since Reock and the Convex Hull method are similar in principle.  In addition to compactness, there are federal laws such as one person, one vote, which ensures each district represents the same number of people. However, after the recent \emph{Louisiana v. Callais}, we do not consider the Voting Rights Act protections in the current work.

\begin{figure}[htbp]
    \centering
    \includegraphics[width=0.45\textwidth]{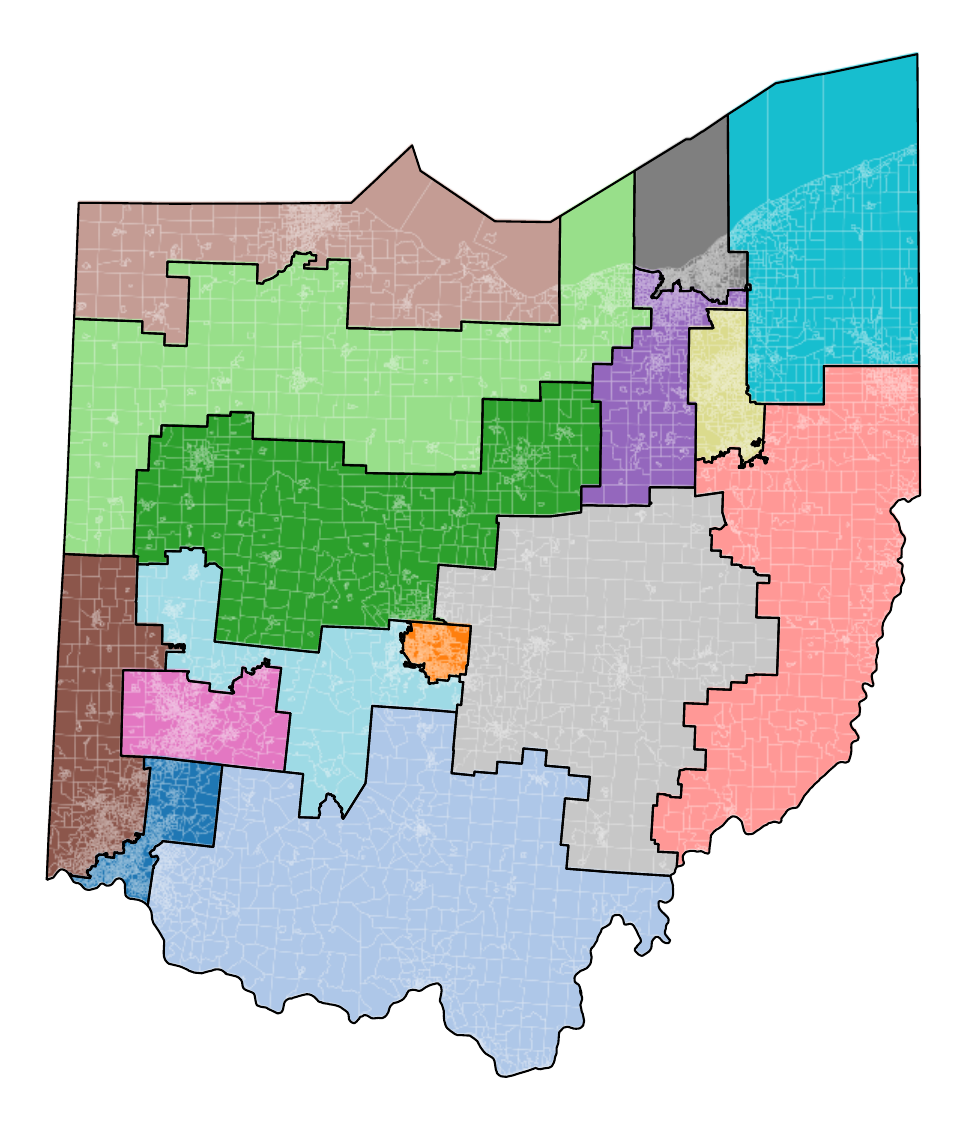}
    \caption{Congressional districts and precincts from the 2022 Remedial plan, with data from VEST project \cite{vest_2020}.}
    \label{fig:2022-precinct-graph}
\end{figure}

When segmenting the state into districts, the Ohio commission typically keeps voting precincts intact. A precinct is a local geographic unit used to administer elections. Each precinct is assigned a polling place, although voters may instead cast ballots early or by mail; those ballots are generally attributed to the voter’s precinct for reporting purposes.
Precincts are local administrative units, and thus are subject to continual updates and changes. They are reported to the nearest census block unit as Voting Districts or VTDs during every census. In the current work, we use the 2020 VEST data which starts with the US Census VTDs and adjusts the boundaries based on voter addresses and reported precinct registration \cite{vest_2020}. Finding exact precinct lines is difficult because Ohio does not aggregate precinct maps at a statewide level, and thus each county administers and stores the precinct boundaries individually. The 2020 VEST Ohio data reports 8,941 Ohio precincts and is displayed in Figure~\ref{fig:2022-precinct-graph}. In the current work, we will consider alternative districting plans that keep precincts whole while allowing for slightly larger population deviations of 2\%; any plan we draw may be refined by splitting precincts to bring the graph into strict population balance with only a small change in population.

\subsection{Ohio Geography and Implications for Redistricting}
In strictly interpreting the congressional redistricting criteria, one must attempt to place districts within cities or townships intersections that have more people than a district, and may not split cities or townships within counties that are between 100,000 people and the size of a congressional district. The 2020 Census estimated that Ohio had a population of 11,799,448 people, meaning that each of the 15 congressional districts should be composed of 786,630 people. In Ohio, only the city of Columbus within Franklin county has more than 786,630 residents, and only two cities have more than 100,000 residents and are in counties that exceed the population of a district: Cleveland and Cincinnati. Akron, Dayton, and Toledo are cities that have more than 100,000 people, however, are not in counties that exceed the population of a district. These three cities are kept together in the three districting plans put forward by the commission.  No township has more than 100,000 people.
Thus, minimally compliant criteria for municipal corporations and townships would need only to account for these three cities, striving to keep a single district within Columbus and strictly keeping the two remaining cities with more than 100,000 people intact. 

However, it is the purview of the redistricting commission to account for additional criteria. For example, the constitution states how to determine if a municipal corporation or township is split, but does not mandate reducing splits:  The commission may decide to reduce splitting these units. Because the Ohio Supreme Court found that that political regions were unduly split in  the 2021 plan, namely in county splitting, it may also be interpreted in the state law to preserve more political regions. Below we will find evidence that the commission did employ such additional criteria.

\subsection{Policy-driven Alternative Plans (or Ensembles) for Auditing}

We have two primary policy questions that will be addressed by a statistical study. The primary policy question asks to what extent the Ohio Redistricting Commission favored a political party. We also ask what consequences these maps had under the 2026 voting patterns.
The first question has been answered, in part, by the courts, but here we frame the question by asking the counterfactual question as to what would have happened had the commission not unduly favored a party.

To address this, we generate an ensemble of plans that reflect the same non-partisan policy considerations as the Ohio redistricting commission. Keeping the non-partisan policy consistent is important in this context: If we were to draw from plans that, say, had far fewer municipal corporation or township splits, then differences in partisan impact between the ensemble of plans and the plans we study may either be due to partisan favor or due to differences in policy. If, however, we build our maps to reflect the non-partisan policy of the maps under audit, then we can be more certain that differences in partisan behavior are due to direct party favoritism, e.g. \cite{herschlag2020quantifying, fifield2020automated, zhao2022mathematically}.

To generate redistricting plans, we first formally define a state space of redistricting plans as a map from precincts to district assignments. We encode the precincts of Ohio as an adjacency graph, $G=(V,E)$, in which nodes represent geographic units, and edges represent adjacency. A redistricting plan is then a population-balanced graph partition with 15 connected regions.

Some precincts, however, are discontiguous and can lead to non-planar connections. We remedy this by connecting the discontiguous components of a precinct by joining the components via the shortest path of intervening precincts within the same county. When several shortest paths are available, we select the one with the smallest total population, breaking ties in favor of the chain with the greatest shared boundary length. We then join the precincts along the path into a single geographic unit (i.e. node in the graph) and update the adjacencies in the graph accordingly. See the Duke Quantifying Gerrymandering group's collection for related papers and reports \cite{duke_quantifying_gerrymandering_papers}, e.g. \cite{chuang_CT_2024}.
The cities of Cleveland and Cincinnati are constitutionally mandated to be unsplit. The cities of Akron, Dayton, and Toledo are all above 100,000 residents and were preserved in all three plans put forward by the commission. Furthermore, it is unclear if the state's Supreme Court would tolerate splitting these regions since it ruled against the 2021 map, in part, due to excessive subdivisions of certain political regions. Thus we also merge the precincts that intersect all five cities.
Under these adjustments, no precinct nor medium sized city will be split. The resulting adjacency graph contains
$|V| = 6,975$
nodes which represent either individual precincts or collections of precincts.

We sample redistricting plans in order to assess the typical district political leaning under a given policy. In sampling an ensemble of redistricting plans, we specify a policy as a probability distribution which must reflect relevant non-partisan policy considerations. The probability may be viewed as a policy because it quantifies a relative preference (or probability) between possible redistricting plans. This is a standard method in auditing plans \cite{herschlag2020quantifying,fifield_automated_2020}.

In the current work, we use the Cycle Walk algorithm to sample graph partitions \cite{deford_cyclewalk_2025}. Cycle Walk uses Metropolis-Hastings on a lifted state space in which districts are represented by trees rather than through a graph partition. Thus we lift our state space to be a balanced-forest. We will weight the edges of the graph to encourage trees with particular community preservation properties. In the framework of the Cycle-Walk algorithm we sample balanced forests with a probability proportional to the product of the edge weights. We let $\xi:V\rightarrow \{1,..., 15\}$ define the partition, and $F$ define a corresponding spanning forest such that $u,v\in V$ are in the same tree if and only if $\xi(u) = \xi(v)$. We let $\xi(F)$ denote the partition induced by $F$. Letting $\pi$ be the probability distribution over forests, we have
\begin{align}
\pi(F)\propto \exp(-J(\xi(F))) \prod_{e\in F} w(e), \quad
\pi(\xi)\propto \exp(-J(\xi)) \prod_{i=1}^{15} \tau(\xi_i),
\end{align}
where $w(e)$ is the weight of edge $e$, $J$ is a score function on partitions (lower scores leading to more probable plans), and $\xi(F)$ is the partition induced by $F$. The second equality, $\pi(\xi)$ represents the probability of seeing the partition event $\xi$, which is a sum over all forests that induce $\xi$; in this equation, $\tau(\xi_i)$ represents the sum of the product of edge weights over all trees in the $i$th region of partition $\xi$.
While $J$ is a policy about the partition, the edge weights may also be thought of as part of the policy: First, redistricting plans with greater total spanning-forest weight tend to correlate with certain measures of compactness \cite{autry_metropolized_2023,deford_recombination_2021}; Second, edge weights provide a locally modulating preference for which adjacent nodes should, or should not, be kept together in a redistricting plan; this has the potential added benefit of aligning proposals in the Markov Chain with the measure.

\subsection{Minimally Compliant Criteria}
Cities with over 100,000 people but less than the number of people per congressional district are automatically preserved. The city of Columbus within Franklin County, however, is made up of more people than can fit in a single district; we also note that Franklin County holds less than two congressional districts and therefore may intersect only two districts. One of these districts must include a ``significant" portion of Columbus (within Franklin): We may measure this by finding the district with the highest population of Columbus residents. An equivalent measure is to examine how many residents of Columbus are \emph{not} in Columbus's best represented district; we denote these people as \emph{Ousted}. Any redistricting plan in Ohio \emph{must} oust Columbus-Franklin residents from the most significant district, since the city within the county is larger than a district; however, counting those outside of the district gives a consistent measure of how well the best representative district was able to incorporate the city's population.

For all other cities in Ohio, a similar methodology works: Anyone not in the district containing the largest share of the population of a municipal corporation or township, is considered to be ousted from that community in the redistricting process. Formally, to evaluate the ousted population of any Ohio municipal corporation or township, $m$, we let $p_{md}$ denote the population of the portion assigned to district $d$, and $p_m = \sum_d p_{md}$, be the total population of the $m$. Then the ousted population is
$$\text{pop}_{\text{ousted}}(M) = \sum_{m\in M}\left(p_m - \max_d p_{md}\right)\,,$$
where $M$ is the set of the municipalities or townships that we wish to consider.
When reporting $\text{pop}_{\text{ousted}}$, we often divide it by the total population of the considered regions and discuss the fraction of considered population that has been ousted.

To determine minimally compliant criteria in Ohio, we begin by modifying $J$ and $w(e)$ in two ways
\begin{enumerate}
    \item For edges with both nodes intersecting the city of Columbus and within Franklin County (where the bulk of Columbus resides), we increase the edge weight in the graph so that $w=5$. This ensures that trees drawn in the space will be less likely to meander in and out of the city.
    \item We take $J= 5 \text{pop}_{\text{ousted}}(C)$ where $C$ represents the set with only the element of Columbus in Franklin. Larger values of $J$ lead to plans with lower probability. 
\end{enumerate}
We have explored just using the weight independently without an ousted population score. This leads to trees that do not often cross the Columbus border, however, there is no stipulation about \emph{where} to cut the resulting trees and thus modulating the edge weight does not control the ousted population. We find that using both modifications together leads to good convergence behavior and ability to tune the amount of observed preservation of Columbus.

Duchin and Spencer tested county preservation in ReCom by changing the random edge weights used to draw minimum spanning trees \cite{duchin_spencer_models_2021}. Other work has used or studied related weighting schemes for minimum spanning trees in ReCom \cite{duchin_south_carolina_2022,tapp_parameter_2025,babson_models_2026}. Here we follow the Cycle Walk approach of DeFord et al.~\cite{deford_cyclewalk_2025} and set $w(e) < 1$ if the two nodes of the edge $e$ are in different counties. This penalizes forests, and hence plans, that split more counties, or have more disconnected parts of a district within a county. In the present work, we find weighting in this way allows us to subsample our Markov chains to achieve all of the county constraints required by the Ohio Constitution and Supreme Court. For details on the chosen cross-county weights see \Cref{apdx:runparams}.

We run Cycle-Walk on 4 independent chains with the above parameters. As stated above, we subsample our ensemble to ensure that Ohio's county splitting criteria is fully satisfied. We test for convergence using the rank-normalized split Gelman-Rubin statistic \cite{gelman_inference_1992, vehtari_rank_2021} on several observables of interest and find good convergence properties, with a maximum and average $\hat{R}$ values of 1.0011 and 1.0004, respectively across our tested observables (see \Cref{apdx:convergence} for more details).

\begin{figure}[htbp]
    \centering
    \includegraphics[width=0.45\textwidth]{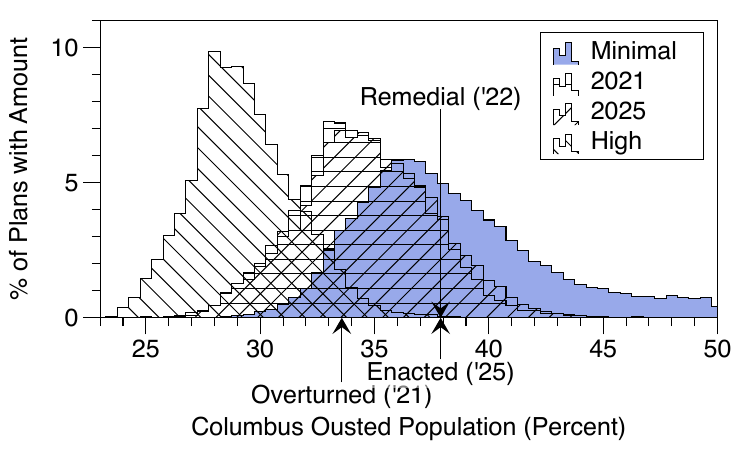}
    \caption{We compare four histograms from our four ensembles that represent the fraction of Columbus's population within Franklin County that has been ousted from the most representative districts. We compare the histograms with the ousted population fraction of the three plans put forth by Ohio legislature.}
    \label{fig:columbus-only-ousted-pop}
\end{figure}

We next compare the non-partisan properties of this ensemble to the properties of the three plans from the legislature. We begin by showing the ousted population of Columbus in \Cref{fig:columbus-only-ousted-pop}: We compare the density of the ousted population in the ensemble (blue) with that of the three plans put forward by the Ohio legislature. We find that the 2022 and 2025 plans oust slightly more of Columbus's population than the average plan in the ensemble. In contrast, the 2021 plan preserves Columbus to a greater degree, however, many plans in the ensemble still preserve Columbus as well or better than the 2021 plan.

\begin{figure}[htbp]
    \centering
    \includegraphics[width=0.45\textwidth]{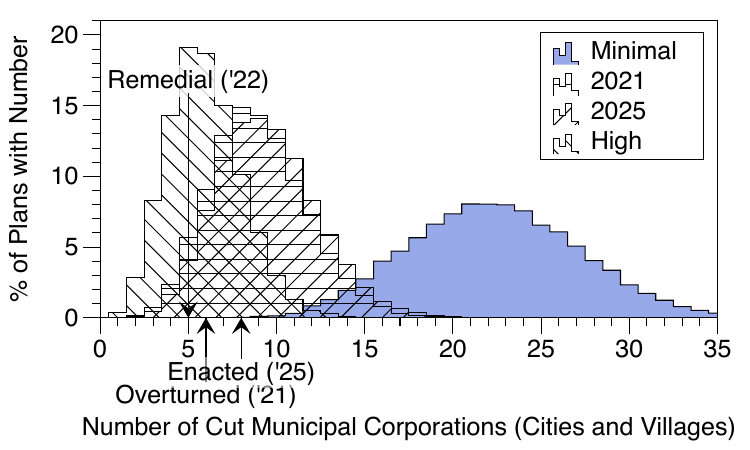}
    \includegraphics[width=0.45\textwidth]{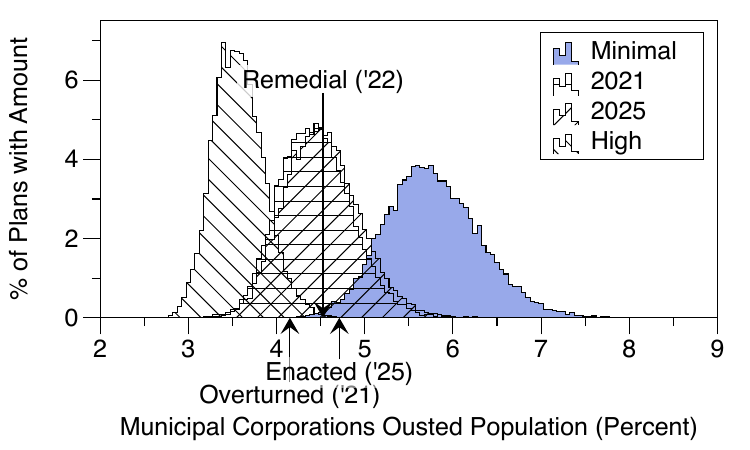} \\
    \includegraphics[width=0.45\textwidth]{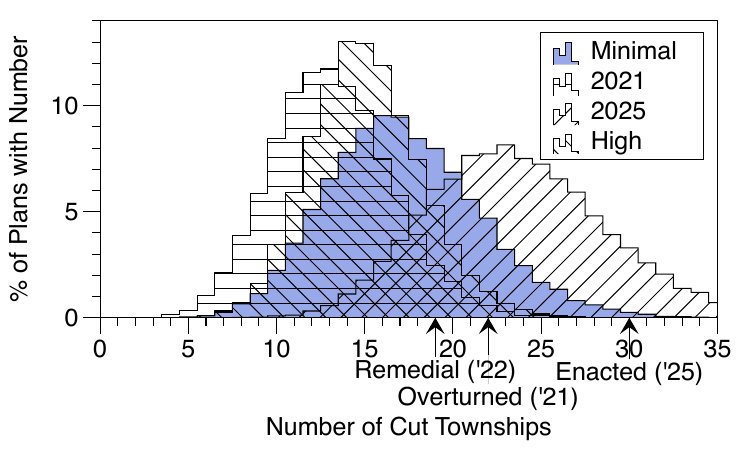}
    \includegraphics[width=0.45\textwidth]{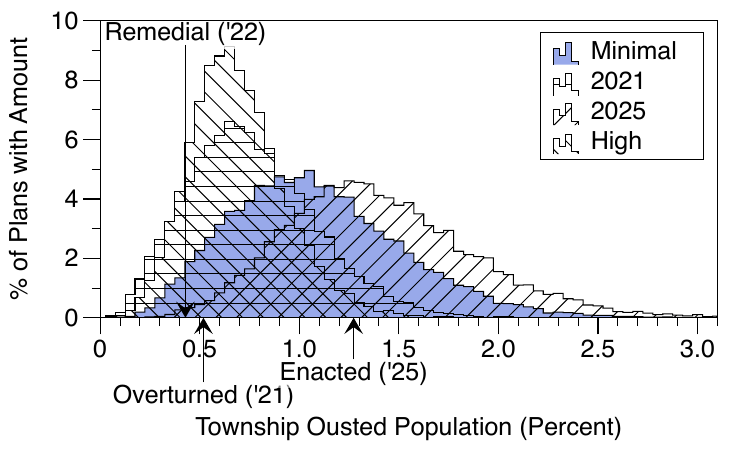} \\
    \caption{We show the number of splits and ousted population of municipal corporations and townships. The number of splits are shown in the left column and the ousted population in the right. The municipal corporation data is show in the top row and the township data on the bottom. The numbers are aggrate across the entire state. Ousted population is shown as the percent of the population within all municipal corporations and townships that have been ousted from the district that best represents them. We show results for all four ensembles we consider and compare with the three plans from the legislature.}
    \label{fig:cuts-and-ousted}
\end{figure}

However, the ensemble was tuned to show similar properties in Columbus's ousted population within Franklin County. When examining the number of split municipal corporations, we see that this ensemble cuts at least 10 municipal corporations, whereas the three plans from the legislature cut at most eight (see top left of \Cref{fig:cuts-and-ousted}).
We also examine the percent of the population in municipal corporations that have been removed from their best represented district (see top right of \Cref{fig:cuts-and-ousted}), and again find that the ensemble ousts a higher fraction of people in municipal corporations than is typically seen in the three plans approved by the legislature.
Thus there is evidence that the commission accounted for keeping municipal corporations intact across the entire state rather than solely in Columbus within Franklin County.

We also examine the number of cuts and ousted population in townships (bottom of \Cref{fig:cuts-and-ousted}). Here we find no evidence that the commission sought to preserve townships in the 2025 plan. In the 2021 and 2022 plans, the number of cut townships are consistent with the ensemble, but the ousted percentage is lower than what is typical of the ensemble. Therefore, we find some evidence that the earlier plans did attempt to preserve townships, even if this evidence is absent for the 2025 plan.

\begin{figure}[htbp]
    \centering
        \includegraphics[width=0.45\linewidth]{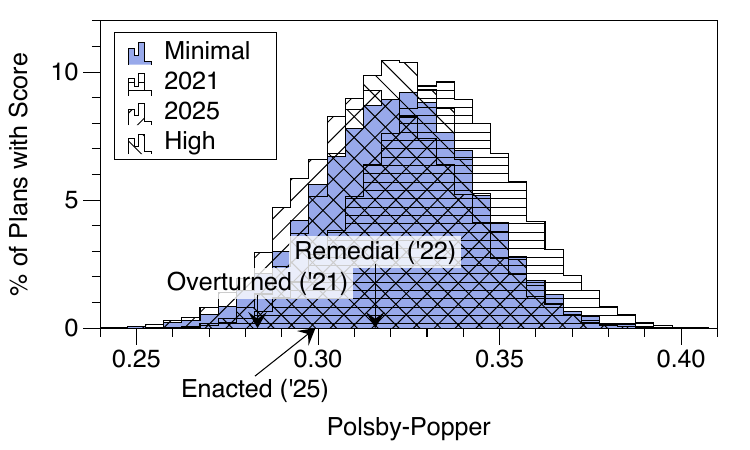}
        \includegraphics[width=0.45\linewidth]{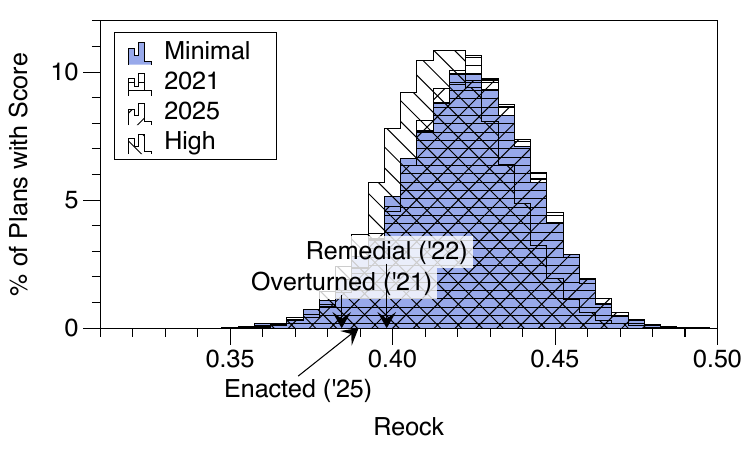}

    \caption{Distributions of compactness scores among the four ensembles specified in
    Table~\ref{tab:ensemble-configurations}, compared with the
    three official plans. We show Polsby--Popper scores (left) and Reock scores (right).}
    \label{fig:compactness}
\end{figure}

We next analyze the compactness properties of the ensemble in comparison to the three plans. We display our results in \Cref{fig:compactness}. We examine both the average Polsby-Popper (left) and Reock (right) scores of each plan. Higher scores mean more compact plans. We find that the three plans under consideration have similar compactness properties to the ensemble.

We conclude that the minimal criteria take many of the non-partisan criteria into account, but not all of it. We next turn to tuning specific parameters to match the non-partisan criteria of the 2021 and 2022 plans, and the 2025 plan.

\subsection{Interpreting the Ohio Redistricting Commission's Non-Partisan Criteria}

Based on the above observations, we propose adding statewide municipal corporation preservation to the policy (probability) function when matching the 2025 plan. We propose adding both municipal corporation and township considerations when matching the 2021 and 2022 plans. Finally, because the courts found that political regions were overly split, we also create an ensemble which seeks better preservation of both municipal corporations and townships.

After tuning our parameters, we arrive at the parameter set presented in \Cref{tab:ensemble-configurations}. Each ensemble also downweights cross-county edges as described further in \Cref{apdx:runparams}. \footnote{The original scripts and commands can be found at \texttt{https://github.com/franklinxzhou/quantifying-gerrymandering-ohio.git}}.

\begin{table}[t]
\centering
\small
\begin{tabular}{llllccc}
\toprule
Ensemble & Municipalities/Townships &
$w_{CV}$ & $\text{pop}_{ousted}^{(CV)}$ & $w_{T}$ &
$\text{pop}_{ousted}^{(T)}$ \\
\midrule
Minimal Criteria   & Columbus in Franklin & 5 & 5 & 1 & 0 \\
Tuned to 2025   & All Cities \& Villages        & 10 & 10  & 1 & 0  \\
Tuned to 2021/22   & All Cities, Villages, \& Townships        & 10 & 10  & 5 & 0  \\
High Preservation   & All Cities, Villages, \& Townships        & 20 & 20  & 5 & 5  \\
\bottomrule
\end{tabular}
\caption{Sampling configurations used in the main analysis. We list which municipal corporations or townships are considered and how edge weights are reweighted between them and how ousted population is accounted for in the energy. Weights/functions with $CV$ only consider cities and villages; those with $T$ only consider townships.}
\label{tab:ensemble-configurations}
\end{table}

We again test for convergence by running 4 independent chains (see \Cref{apdx:convergence}) and compare the outsted population within Columbus \Cref{fig:columbus-only-ousted-pop}, across all municipal corporations and townships \Cref{fig:cuts-and-ousted} (right side; top and bottom, respectively), the number of cut municipalities and townships \Cref{fig:cuts-and-ousted} (left side; top and bottom, respectively), and the compactness scores \Cref{fig:compactness}. The ensembles are shown with clear patterns with horizontal lines (2021/2022 criteria), upward sloping lines from left to right (2025 criteria) and downward sloping lines (high preservation of political region criteria) in all plots.

We find good agreement between the three plans and the ensembles adhering 2021/2022 criteria and 2025 criteria in the Columbus ousted population, the statewide cuts and ousted population for municipal corporations, and compactness. For township criteria, we find good agreement between the ousted population when matching the derived ensembles to the corresponding plans: The 2021 and 2022 plans reduce ousted population in statewide townships to a far greater degree than the 2025 plan.

The ensemble that highly preserves political regions performs similarly or better to all of the other ensembles when considering cut counts and ousted population and similarly with regards to compactness. This ensemble shows that the commission had many opportunities to further preserve political geography.

\section{Classifying the Extent of Party Favoritism}

We now turn to the electoral consequences of the three official plans. To account for the potential influence of partisan electoral outcomes in action, we use the ensemble calibrated to match the level of municipal preservation of the Enacted (2025) plan as our reference ensemble. Using other ensembles as the reference ensemble uncovers similar electoral outcomes, as shown in \Cref{apdx:electoral-analysis-other}. We examine whether the electoral outcomes under the official plans are consistent with those observed among plans drawn under comparable districting criteria.

\begin{table}[t]
\centering
\caption{Statewide elections used in the electoral analysis. How we estimate the precinct-level vote share in November 2026 elections is discussed in Appendix~\ref{apdx:2026}.}
\label{tab:elections}
\begin{tabular}{ll}
\toprule
Year & Contests \\
\midrule
2016\cite{vest_2016} & President; U.S. Senate \\
2018\cite{vest_2018} & U.S. Senate; Governor; Attorney General; Auditor; \\
     & Secretary of State; Treasurer \\
2020\cite{vest_2020} & President \\
2022\cite{rdu_2022} & U.S. Senate; Governor; Attorney General; Auditor; \\
     & Secretary of State; Treasurer; Chief Justice; \\
     & Supreme Court (terms commencing January 1 and January 2) \\
2024\cite{rdh_2024} & President; U.S. Senate; Supreme Court (terms commencing \\
     & January 1 and January 2); Supreme Court unexpired term \\
2026 & U.S. Senate; Governor; Attorney General; Auditor; \\
& Secretary of State; Treasurer; \\
& Supreme Court (terms commencing January 1 and January 2)\\
\bottomrule
\end{tabular}
\end{table}

\begin{figure}[htbp]
    \centering
    \includegraphics[width=0.99\textwidth]{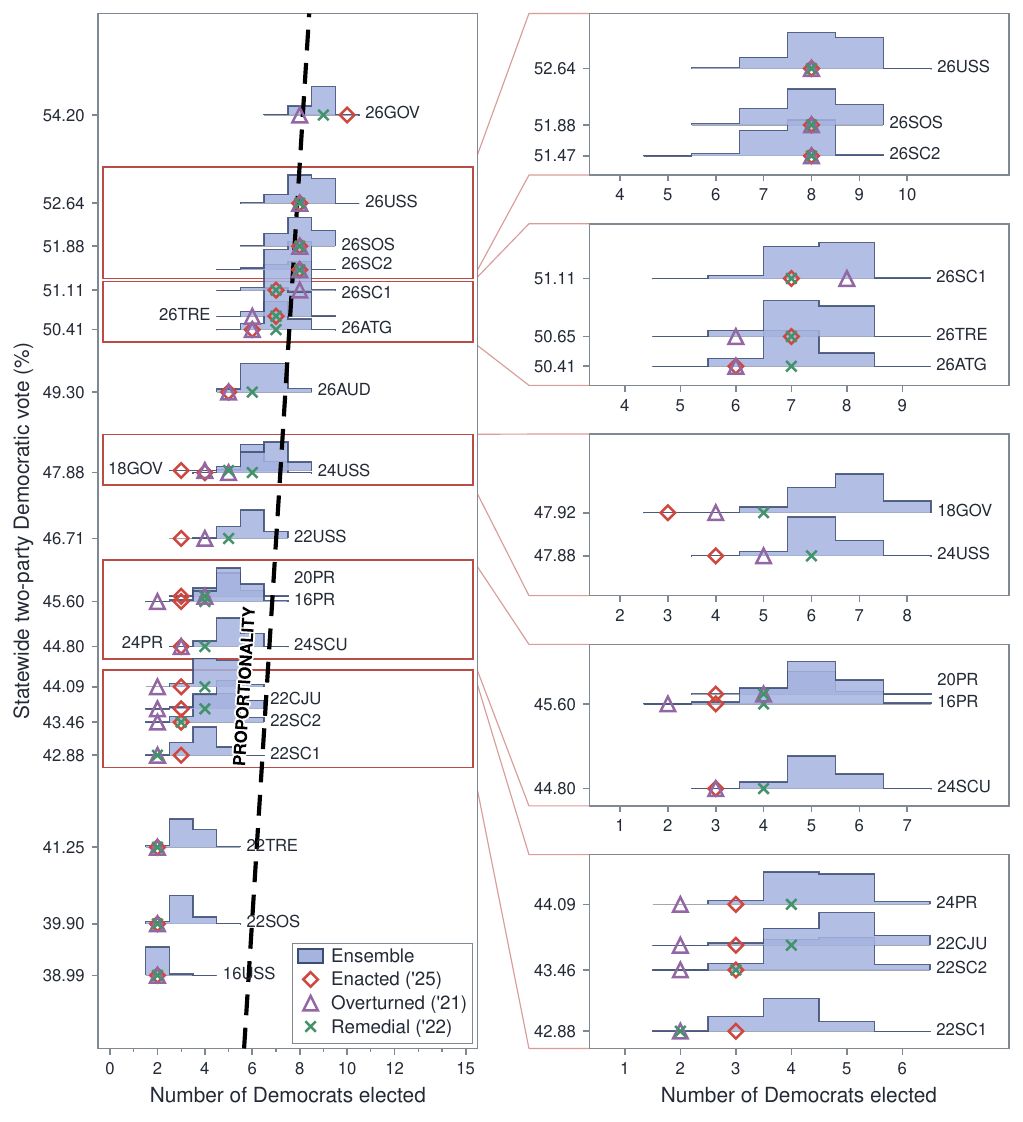}
    \caption{Collected Seat Histogram of the number of Democratic seats in the ensemble that matches 2025 enacted plan in municipal preservation, based on selected historical statewide elections in Table~\ref{tab:elections}. In the Y-axis, the historical elections are ordered according to the Democratic vote share. The colored dots on the X-axis represent the number of Democratic seats based on official plans. The dashed lines represent the Democratic seat counts based on proportional representation.}
    \label{fig:ceah-zoom}

    \vspace{0.5em}
    \begin{minipage}{\textwidth}
        \footnotesize
        \textit{Note:} Election labels consist of a two-digit election year followed by an abbreviation for the office: PR (President), USS (U.S. Senate), GOV (Governor), ATG (Attorney General), AUD (Auditor of State), SOS (Secretary of State), TRE (Treasurer of State), and CJU (Chief Justice of the Ohio Supreme Court). SC1 and SC2 denote Ohio Supreme Court elections for distinct full terms, while SCU denotes an election for an unexpired term.
    \end{minipage}
\end{figure}

\subsection{Suppressed Democratic Representation}
Under the same statewide voting patterns, Ohio's official plans from 2021 and 2025 generally yield fewer Democratic seats than plans in the reference ensemble when Democrats receive roughly 42--48\% of the statewide two-party vote. To make this comparison, we apply the statewide election returns in Table~\ref{tab:elections} to each plan. We hold precinct vote totals fixed, aggregate them within each plan's congressional districts, and count the districts in which Democrats receive more than half of the two-party vote. Figure~\ref{fig:ceah-zoom} shows the distribution of these counterfactual seat counts in the ensemble for each election, alongside the counts for the three official plans.

For example, Democrats received 47.92\% of the two-party vote in the 2018 gubernatorial election. Applying those returns gives Democrats three seats under the 2025 plan, four under the 2021 plan, and five under the 2022 plan. Sampled plans typically yield five to eight Democratic seats, with seven the most common outcome. On average, the ensemble yields fewer Democratic seats than proportionality would imply.

Congressional districts provide localized representation. The ensemble shows how Ohio's voting geography and the modeled non-partisan criteria together translate statewide votes into seats. Compared with that baseline, the official plans, particularly those from 2021 and 2025, generally deepen the Democratic seat shortfall under statewide Democratic vote shares between 42-48\%.

\begin{figure}[htbp]
    \centering

    \begin{subfigure}[t]{0.49\textwidth}
        \centering
        \includegraphics[width=\linewidth]{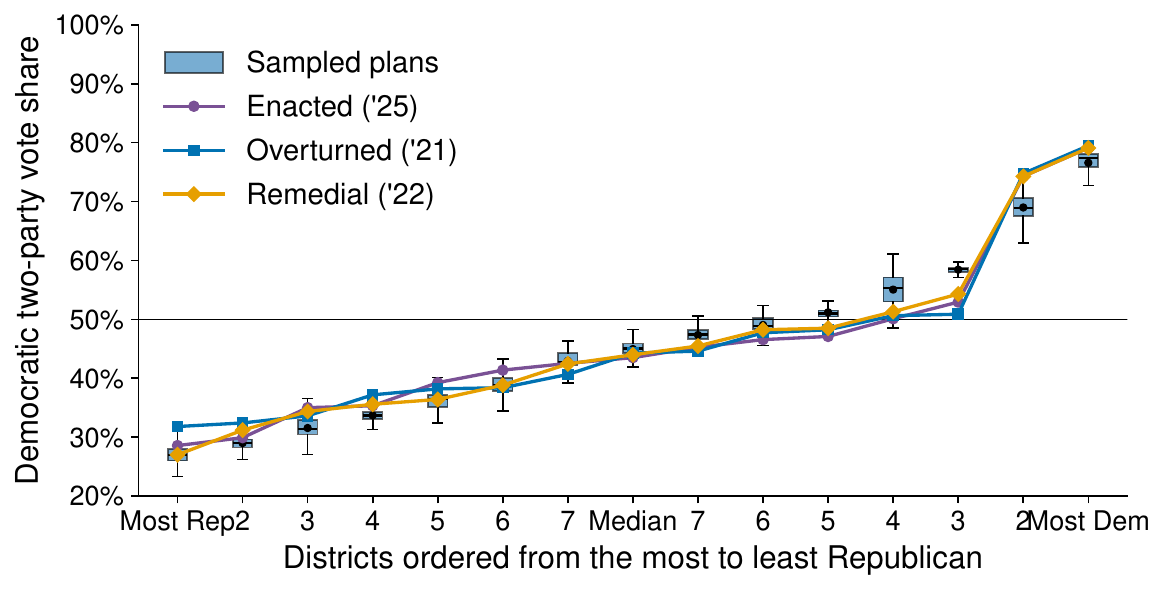}
        \caption{2020 presidential election, ensemble.}
        \label{fig:20-pre}
    \end{subfigure}
    \hfill
    \begin{subfigure}[t]{0.49\textwidth}
        \centering
        \includegraphics[width=\linewidth]{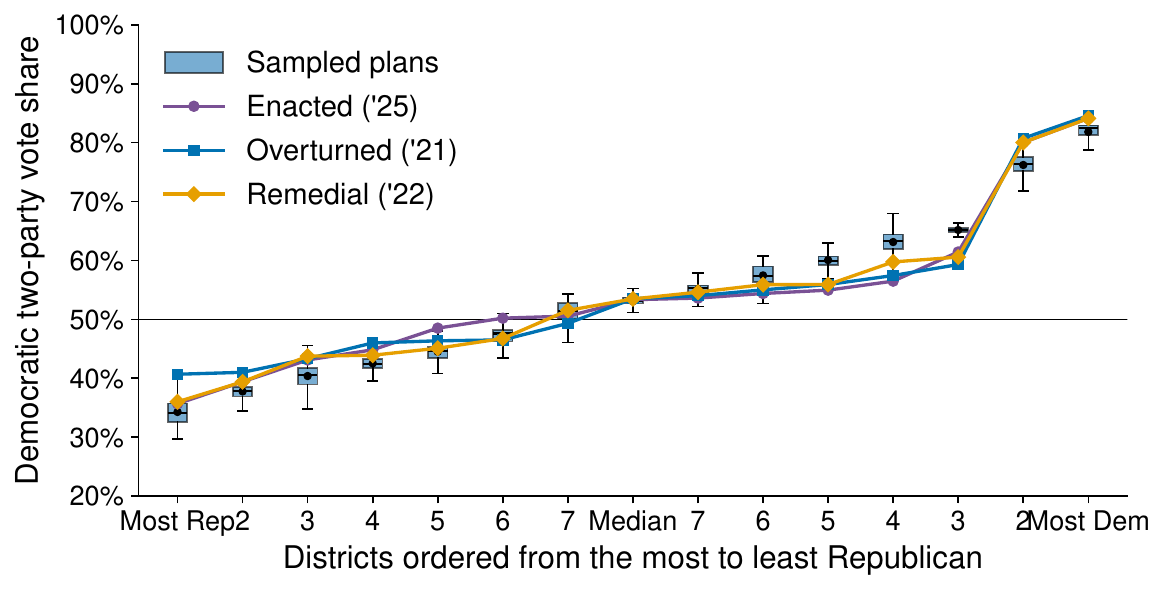}
        \caption{2026 gubernatorial election, ensemble.}
        \label{fig:26-gov}
    \end{subfigure}

    \caption{We display the Democratic two-party vote shares by district rank in Ohio congressional plans. We rank districts within each plan from most Republican to most Democratic. At each rank, the boxplots show the distribution among plans in the ensemble tuned to the 2025 enacted plan's municipal preservation.
    The colored lines trace the three official plans, and the horizontal line marks 50\%; above and below this line represents a Democratic or Republican leaning district, respectively.}
    \label{fig:ordered-partisan-profiles}
\end{figure}

Seeing \emph{who} (or which party) wins over a collection of elections is revealing; however, it is also important to consider \emph{how much} they win by. To examine this we order the districts within each plan by their \emph{partisan lean} which we define to be the fraction of Democratic votes of the Democratic and Republican votes (not considering third party votes; see \Cref{fig:20-pre,fig:26-gov}). In all three official plans, the two most Democratic districts have higher Democratic vote shares than those in 90--95\% of sampled plans, depending on the election considered, suggesting that Democrats may have been overly concentrated in these two districts.

We next measure Republican vote concentration in the third-most through fifth-most Democratic districts. For each plan and election, we sum the Republican votes in those three districts and divide by all Republican votes cast statewide in that election. For each official plan, fewer than 0.1\% of ensemble plans had a larger share in at least seven of the 31 elections, and at most 1\% had a larger share in at least 21 elections. This pattern is consistent with cracking. The fifth-most Democratic district in the 2020 presidential election illustrates the consequence for representation. Its ensemble median is just above 50\% Democratic, and all three official plans place it below 50\%.

Applying this statistic of Republican vote concentration in the third-most through fifth-most Democratic districts provides further evidence of the intent behind the maps. For each map, the Republican concentration in the third-most through fifth-most Democratic districts is significantly higher than the ensemble for nearly every election that the mapmakers had access to when drawing the map.
The share of statewide Republican voters in the third-most through fifth-most Democratic districts is greater in the 2021 map than it is in 98\% of plans in the ensemble for every pre-2021 election other than a 2016 and 2018 race.
Similarly, the share of statewide Republican voters in the third-most through fifth-most Democratic districts for the 2022 map is greater than it is in 98\% of plans in the ensemble for every pre-2022 election (and greater than 99\% of the ensemble plans for all but two 2018 elections).
For the 2025 map, the share of statewide Republican voters in the third-most through fifth-most Democratic districts is greater than 99\% of the ensemble in every pre-2026 election.

For the 2021 and 2022 plans, these findings are consistent with the Ohio Supreme Court's rulings on Republican favoritism \cite{adams_v_dewine_2022,neiman_v_larose_2022}. We find the same patterns in the 2025 plan.

\subsection{The Risks of Gerrymandering}

Applying the estimated precinct returns from the 2026 gubernatorial election, in which Democrats received 54.20\% of the statewide two-party vote, yields ten Democratic seats under the 2025 plan. The most common outcome in the ensemble is nine, and strict proportionality corresponds to 8.13 seats (\Cref{fig:ceah-zoom}). Even in this race, the third-most through fifth-most Democratic districts have lower Democratic vote shares under the 2025 plan than the ensemble medians at those ranks (\Cref{fig:26-gov}).

The tenth seat comes from the sixth-most Republican district in the 2025 plan. Its Republican share of the two-party vote is low relative to the ensemble at that rank, where the median district still leans Republican. Under the 2025 plan, Democrats narrowly cross 50\%. This comparison illustrates how a relatively low Republican vote share at this rank may produce an additional Democratic seat.

In several other 2026 races, the official plans yield seat counts closer to the center of the ensemble. For example, applying the estimated U.S. Senate returns at 52.64\% statewide Democratic support gives all three official plans eight Democratic seats, the most common ensemble outcome.

The possibility that Democrats can overcome this firewall should not be interpreted as evidence that the underlying maps are fair or that they do not unduly provide partisan favoritism. Under election structures pre-2026, Republicans gain an undue advantage. However, the 2025 map shows that under the 2026 elections, it is possible for Democrats to win additional seats beyond what is typical of the ensemble and in a way that departs further from proportionality and favors Democrats.

\subsection{Partisan Asymmetry}

Proportionality is a common, but not mandated concept to understand partisan favoritism. Another common measure is partisan symmetry, which concerns whether opposite statewide partisan vote shares translate into equivalent levels of representation for the two parties. This was one of the key measures of partisan favoritism considered by the Supreme Court of Ohio \cite{adams_v_dewine_2022, league2022opinion}; the other measures of efficiency gap, mean-median, and declination are derived from various ways of quantifying symmetry. We assess the extent to which Ohio's three official plans and our ensemble plans exhibit partisan symmetry by directly examining symmetry under hypothetical elections. Herschlag and Mattingly used this approach to evaluate proposed remedial congressional and state Senate plans in North Carolina, as discussed in the Harper plaintiffs' 2022 response in \textit{Harper v.\ Hall} \cite{harper_plaintiffs_remedial_response_2022}. We then compare the number of seats awarded to the two parties under reciprocal statewide election conditions.

\begin{figure}[htbp]
    \centering

    \begin{subfigure}[t]{0.49\textwidth}
        \centering
        \includegraphics[width=\linewidth]{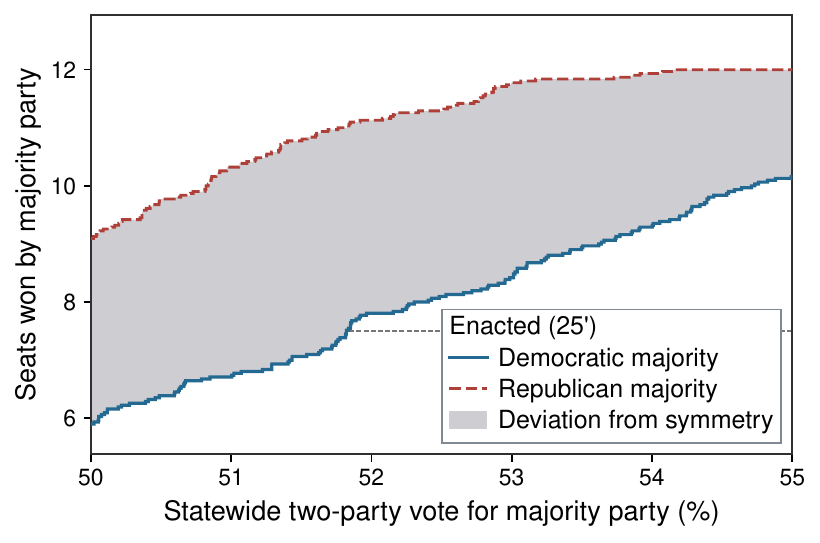}
        \caption{Enacted (2025)}
        \label{fig:asymmetry-enacted}
    \end{subfigure}
    \hfill
    \begin{subfigure}[t]{0.49\textwidth}
        \centering
        \includegraphics[width=\linewidth]{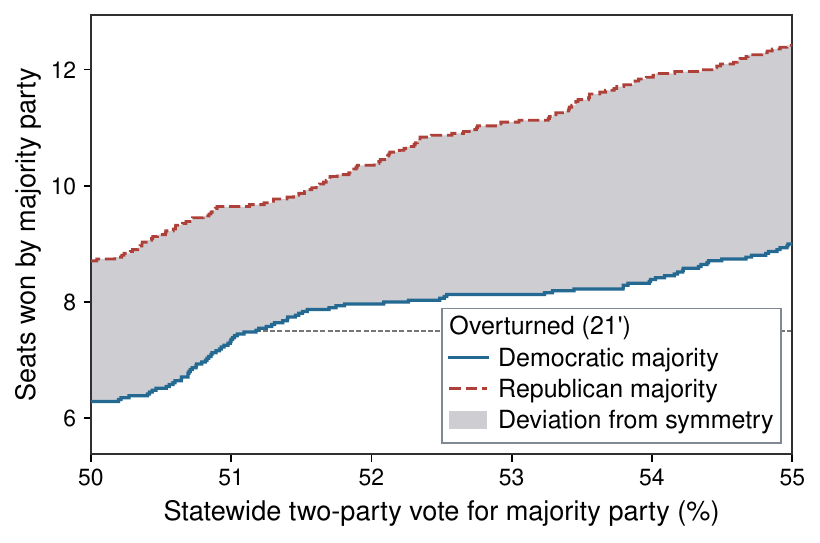}
        \caption{Overturned (2021)}
        \label{fig:asymmetry-overturned}
    \end{subfigure}

    \medskip

    \begin{subfigure}[t]{0.49\textwidth}
        \centering
        \includegraphics[width=\linewidth]{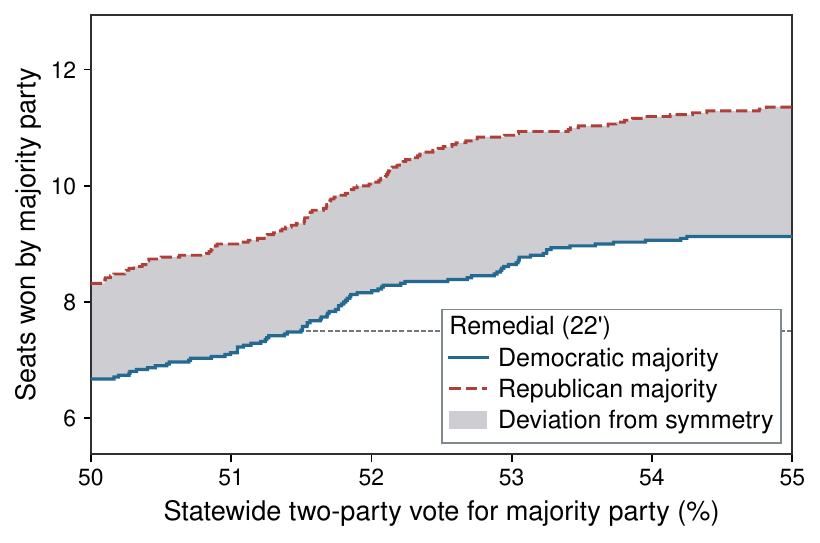}
        \caption{Remedial (2022)}
        \label{fig:asymmetry-remedial}
    \end{subfigure}
    \hfill
    \begin{subfigure}[t]{0.49\textwidth}
        \centering
        \includegraphics[width=\linewidth]{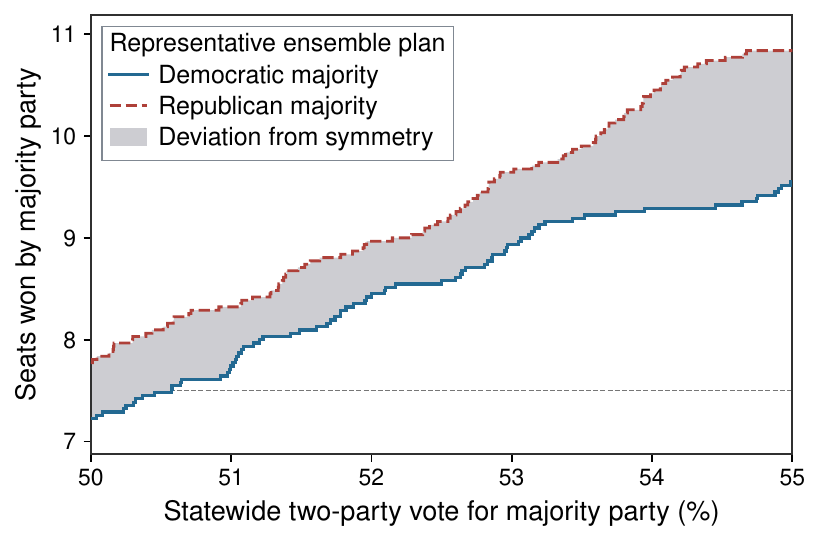}
        \caption{Sampled plan}
        \label{fig:asymmetry-sampled}
    \end{subfigure}

    \caption{We show the statewide vote percentage received by the party winning the popular vote compared with the statewide seats won by the party in the 31 historic elections in Table~\ref{tab:elections} under the three official plans and a sampled plan with a near-median grey area between the two curves in the ensemble that match the municipal preservation of Enacted 2025 plan. The area of the shaded region shows an average deviation from symmetry. In a perfectly symmetric map, the blue line should always agree with the red line.}
    \label{fig:partisan-asymmetry-curves}
\end{figure}

\begin{figure}[htbp]
    \centering
    \includegraphics[width=0.45\textwidth]{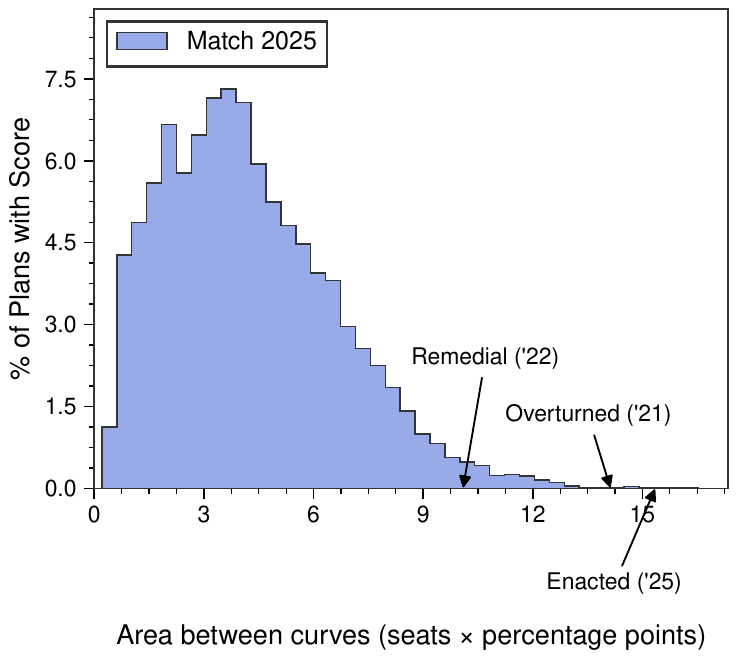}
    \caption{Distribution of partisan asymmetry scores among 20,000 sampled plans under the Tuned to 2025 criteria. The score is defined as the area between the reciprocal seats-votes curves, averaged across elections in Table~\ref{tab:elections}, over the range in which the respective party receives 50-55\% of the statewide two-party vote.}
    \label{fig:asymmetry-histogram}
\end{figure}

To construct such reciprocal electoral conditions, we adjust the Democratic vote share in each district by a fixed amount to achieve a statewide Democratic vote share, a technique commonly known as uniform swing. Then, for each election and plan, we construct a range of statewide vote shares $v\in [0.5, 0.55]$. For each $v$, we compare the number of Democratic seats obtained when Democrats receive a statewide vote share of $v$ with the number of Republican seats obtained when Republicans receive a statewide vote share of $v$. We average the seat counts across all elections in Table~\ref{tab:elections}. In a perfectly symmetric map, the Republican seat count curve should always agree with the Democratic seat count curve, so the area between two curves reflects the extent of partisan asymmetry.

The resulting curves for the three official plans and one representative ensemble plan that has the median area between the two parties' seat count curves among plans in the ensemble matching the municipal preservation of Enacted 2025 plan are displayed in Figure~\ref{fig:partisan-asymmetry-curves}.
Only 4 out of 20,000 plans (0.02\%) in the ensemble are less symmetric than the 2025 plan. For the 2022 plan, only 16 (0.08\%) plans from the ensemble are less symmetric. For the 2021 plan, 383 (1.9\%) plans from the ensemble are less symmetric.
Most plans in the ensemble have a much smaller deviation from partisan symmetry than the three Commission plans; as presented in Figure~\ref{fig:asymmetry-histogram}, all three official plans exhibit a large deviation from partisan symmetry relative to the ensemble, with the Enacted 2025 plan exhibiting the largest deviation among the three.

\section{Improving Municipal Preservation and Partisan Fairness}

We shift our attention to the high preservation ensemble to ask if it would have been possible for the commission to further improve the level of municipal corporation and township preservation and if doing so would have an impact on the typical partisan outcome of the maps. In a past expert report, we have seen evidence that such policy shifts can lead to changes in the partisan make up of ensembles \cite{mattingly_nc_expert_report_2021}.

Above, we have shown that the high preservation ensemble showcases the opportunity to further preserve municipal corporations and townships. \Cref{fig:columbus-only-ousted-pop,fig:cuts-and-ousted} show that the Columbus and statewide municipal corporation ousted populations may be significantly reduced, while still maintaining a low level of township ousted population. These figures also show that the number of both municipal and township cuts may be improved beyond what was achieved by the Ohio Commission. Furthermore, all of this may be achieved while preserving the compactness of the districts, as shown in \Cref{fig:compactness}.

\begin{figure}[htbp]
    \centering

    \begin{subfigure}[t]{0.495\textwidth}
        \centering
        \includegraphics[width=\linewidth]{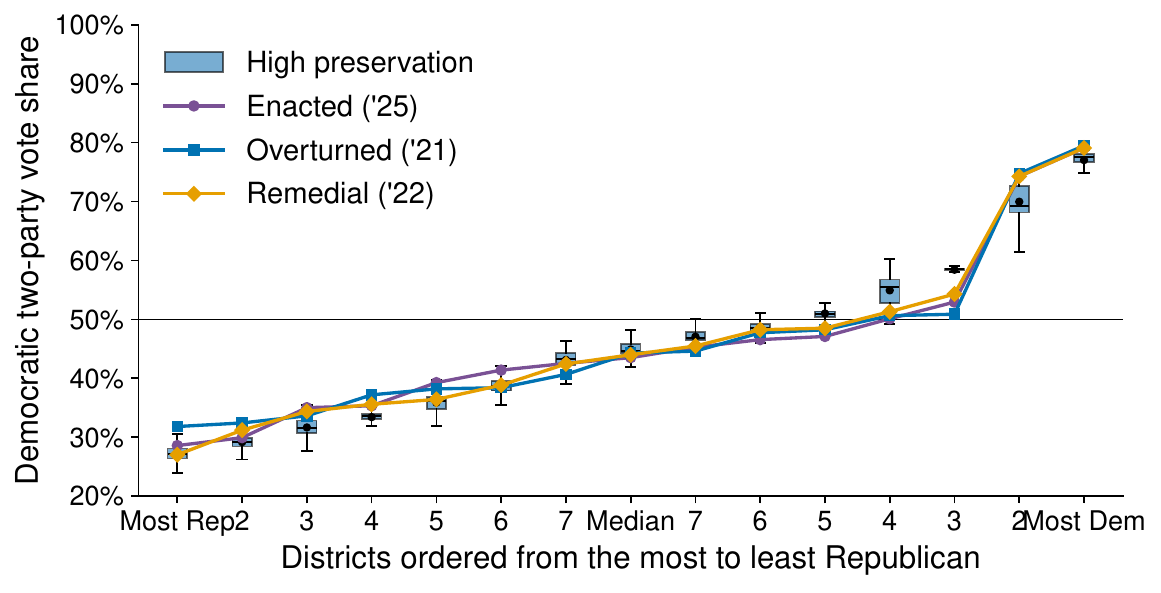}
        \caption{2020 presidential election, ensemble.}
        \label{fig:20-pre-hp}
    \end{subfigure}
    \hfill
    \begin{subfigure}[t]{0.495\textwidth}
        \centering
        \includegraphics[width=\linewidth]{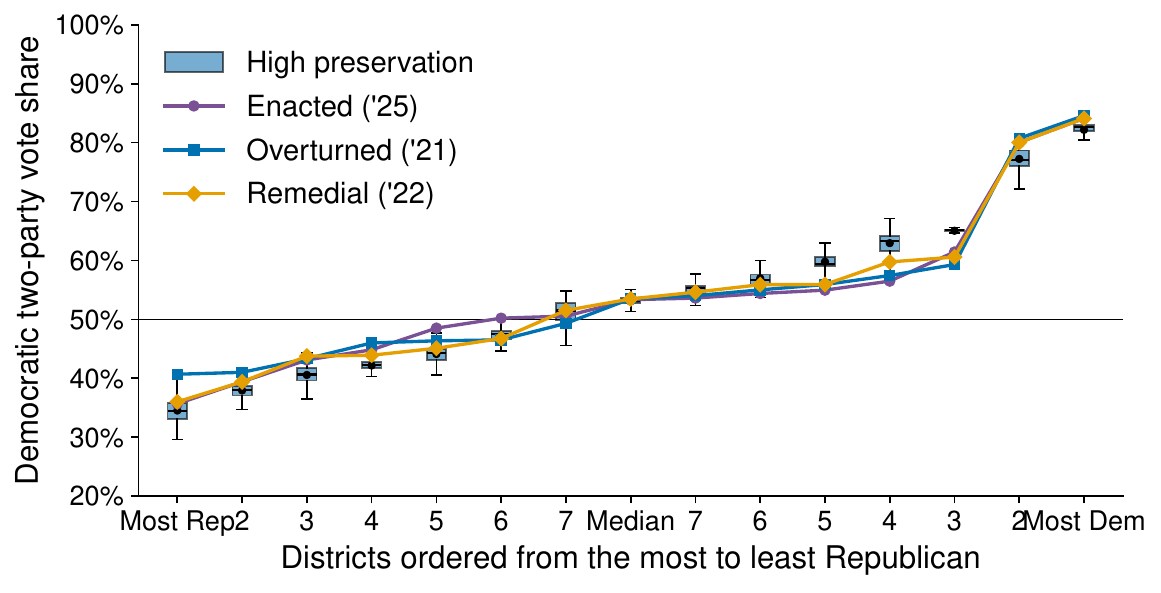}
        \caption{2026 gubernatorial election, ensemble.}
        \label{fig:26-gov-hp}
    \end{subfigure}

    \medskip

    \caption{We display the Democratic two-party vote shares by district rank in Ohio congressional plans. We rank districts within each plan from most Republican to most Democratic. At each rank, the boxplots show the distribution among plans in the high preservation ensemble.
    The colored lines trace the three official plans, and the horizontal line marks 50\%; above and below this line represents a Democratic or Republican leaning district, respectively.}
    \label{fig:ordered-partisan-profiles-hp}
\end{figure}

In Figure~\ref{fig:ordered-partisan-profiles-hp}, again, we compare the high-preservation ensemble with Ohio's three official plans against the seats across the 2020 Presidential and 2026 Gubernatorial elections. Figure~\ref{fig:ordered-partisan-profiles-hp} demonstrates that the Democratic support is atypically concentrated in the two most Democratic districts, while the official plans yield a substantially lower Democratic support in the third-most through fifth-most Democratic districts.
The analysis of other statewide elections and the number of Democratic seats yielded by the official plans can be found in Figure~\ref{fig:ceah-zoom-hp} in Appendix~\ref{apdx:electoral-analysis-other}. The three plans yield fewer Democratic seats than plans in the high preservation ensemble within the Democratic statewide vote share range of 42-48\%. We also examine Figure~\ref{fig:asymmetry-hp}, which demonstrates that, like the Tuned to 2025 ensemble, the plans in the high preservation ensemble are more symmetric than the three official plans. With the two measures of plan fairness, we have observations on electoral patterns consistent with the ones made with the ensembles tuned to match the official plans' municipal preservation policies. We rule out the possibility that unfair electoral results arise from a high level of preservation of municipal corporations and townships.

\section{A Look into Publicly Submitted Plans}

The Ohio Redistricting Commission makes plans submitted by the public available. We extracted 104 districting plans from these submissions. The public drew these plans independently of the official map-drawing process and our sampling procedure. We use the collection for descriptive comparisons with the official plans and the 2025 ensemble.

\begin{figure}[htbp]
    \centering
    \includegraphics[width=0.45\textwidth]{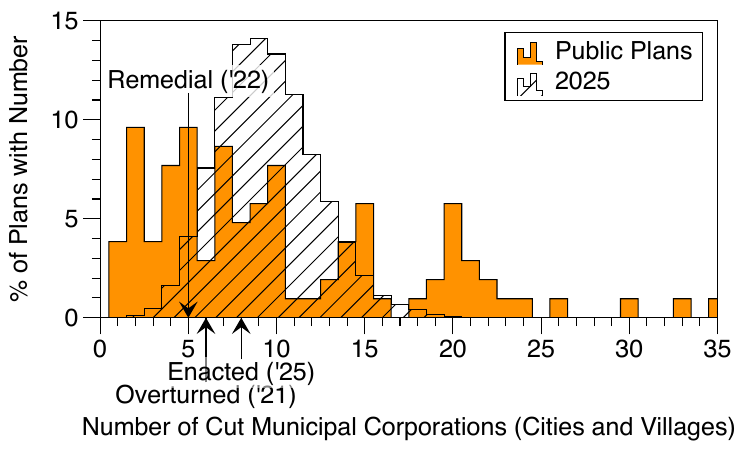}
    \includegraphics[width=0.45\textwidth]{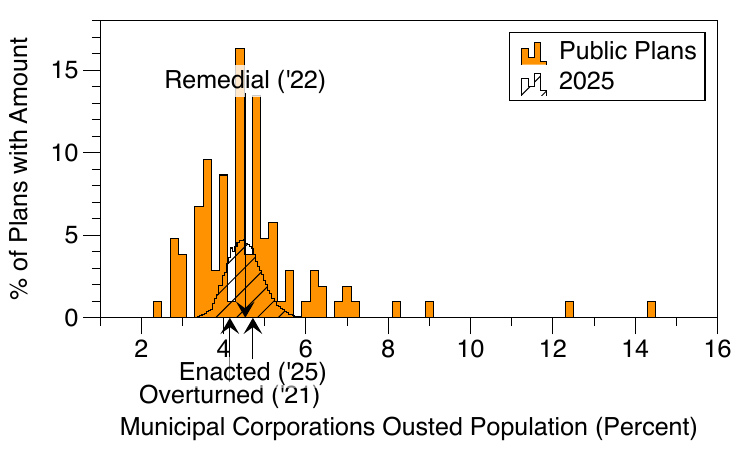} \\
    \includegraphics[width=0.45\textwidth]{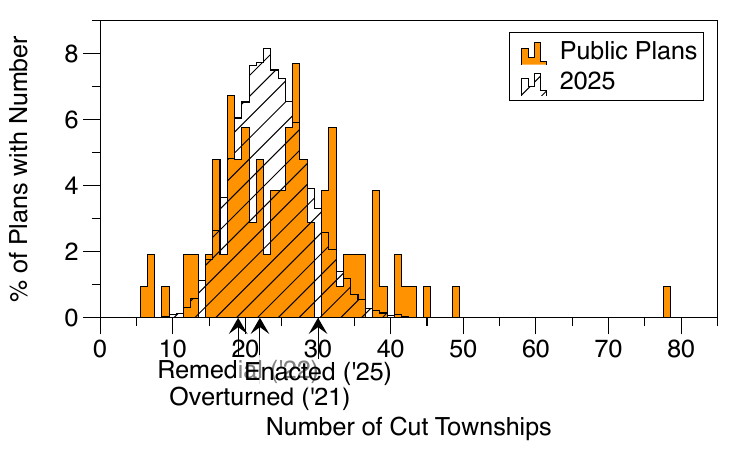}
    \includegraphics[width=0.45\textwidth]{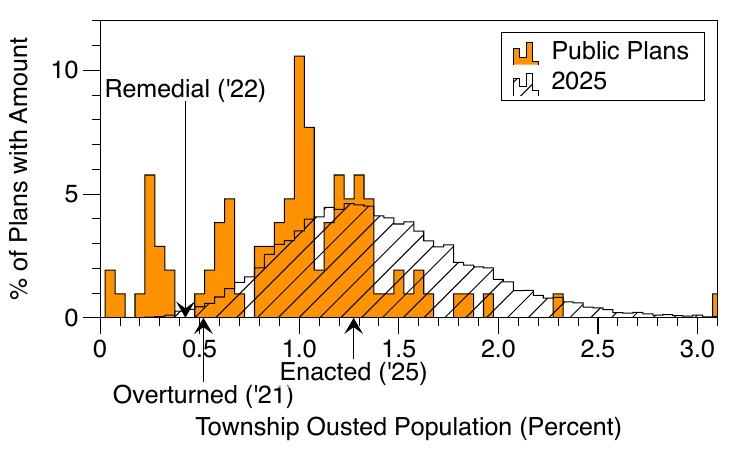} \\
    \caption{The left panels show the number of cut municipal corporations (top) and townships (bottom) statewide. The right panels show the percentage of the population within those units that has been ousted from the district that best represents them. We compare the public plans with municipal-preservation scores with the 2025 ensemble and the three official plans.}
    \label{fig:cuts-and-ousted-public}
\end{figure}

We begin by comparing the non-partisan considerations of the public plans and the ensemble. We examine the municipal corporation and township cutting in \Cref{fig:cuts-and-ousted-public} and compare the properties of the public plans with the 2025 ensemble and the three official plans.
We find that although some plans clearly ignore municipal corporations and townships, many of them have comparable or better preservation of units. When examining the average Polsby-Popper compactness score, however, only 24 of the 104 plans have levels of compactness that are within the support or better than the 2025 ensemble. The majority of the plans, 82 out of 104 plans, comply with the county splitting criteria: 5 of them fail XIX.2(B)(5) (number of split counties) and 23 fail XIX.2(B)(7) (two districts cannot share multiple counties). The 5 that fail the first criterion are amongst the 23 that fail the second. 52 of the 104 public plans split a county into more than two districts; this is not forbidden by the constitution and it is unclear whether the courts would interpret this as overly splitting political regions.

\begin{figure}[htbp]
    \centering
    \begin{subfigure}[t]{0.49\textwidth}
        \centering
        \includegraphics[width=\linewidth]{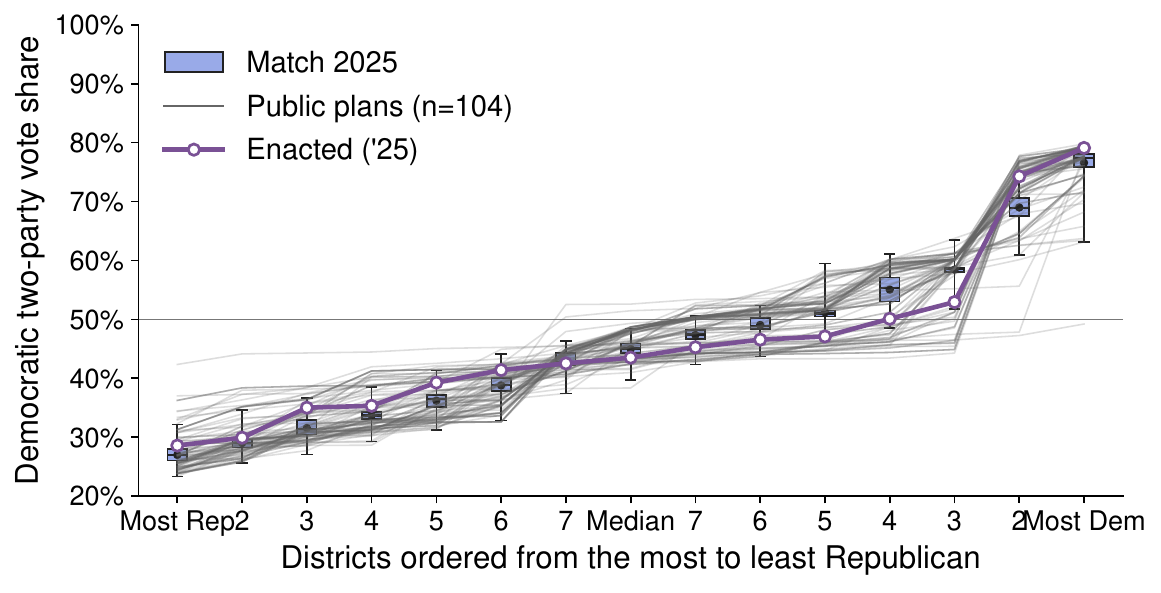}
        \caption{2020 Presidential Election.}
        \label{fig:20-pre-public-vs-ensemble}
    \end{subfigure}
    \begin{subfigure}[t]{0.49\textwidth}
        \centering
        \includegraphics[width=\linewidth]{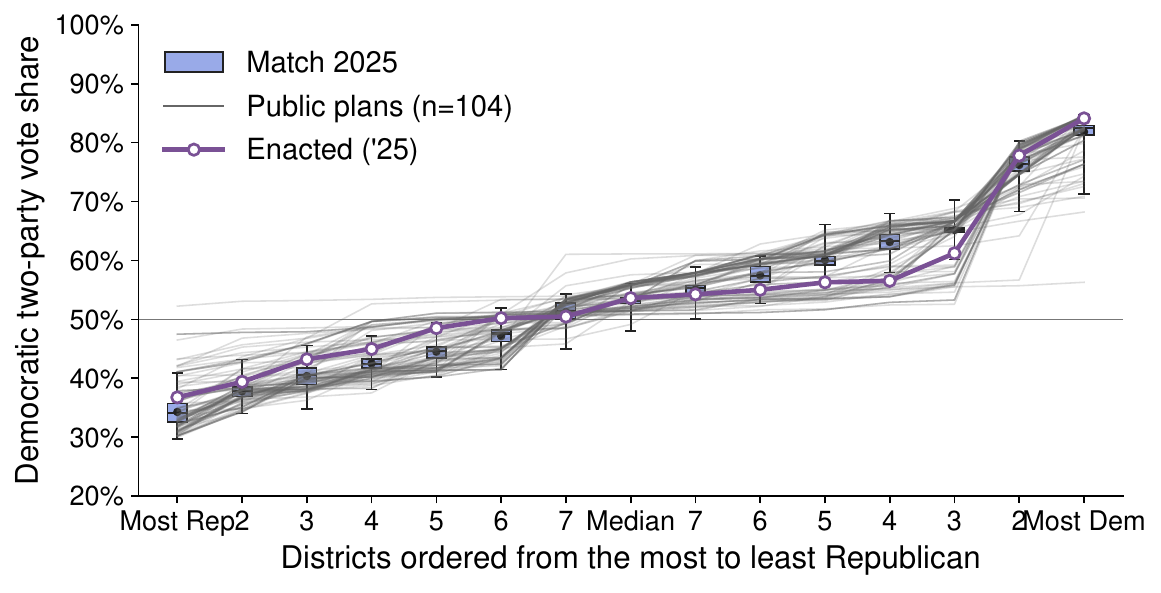}
        \caption{2026 Gubernatorial Election.}
        \label{fig:26-gov-public-vs-ensemble}
    \end{subfigure}

    \begin{subfigure}[t]{0.49\textwidth}
        \centering
        \includegraphics[width=\linewidth]{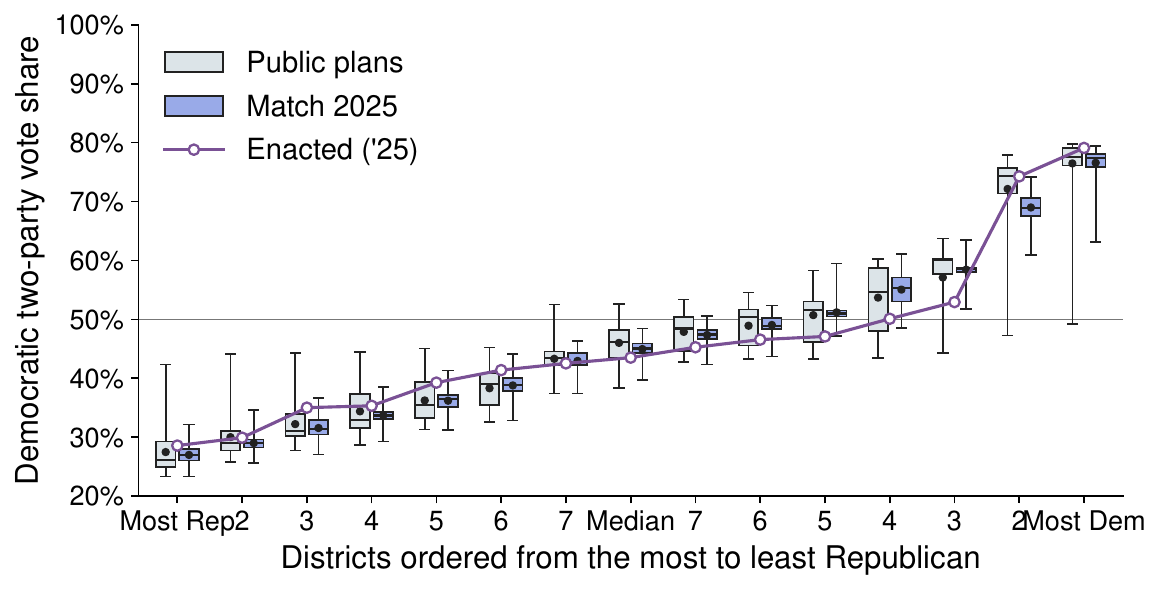}
        \caption{2020 presidential election, comparison by district rank.}
        \label{fig:20-pre-public-vs-ensemble-boxplots}
    \end{subfigure}

    \caption{We display the two-party vote share by district rank for public plans and the ensemble tuned to match the municipal preservation policy of the Enacted 2025 plan. In panels (a) and (b), gray lines trace each plotted public plans, box plots summarize the ensemble at each rank, and the colored line traces the Enacted 2025 plan. The horizontal line marks 50\%. Panel (c) compares the rank-wise distributions of the public plans and the ensemble under the 2020 presidential returns.}
    \label{fig:ordered-partisan-profiles-public}
\end{figure}

We examine the partisan properties of the plans in \Cref{fig:ordered-partisan-profiles-public}. We again order the districts from least to most Democratic. The gray lines show electoral profiles for the public submissions. We compare them with the ordered marginal box plots from the 2025 ensemble and with the Enacted 2025 plan. We see a wide range of outcomes. For example, Carson Spreeman's plan is the only one in which Republicans win all 15 seats under the 2020 presidential returns. Under the 2026 gubernatorial returns, Democrats win all 15 seats in the same plan. Addison Morelle's plan shows a sharp increase in Democratic vote share at the seventh-most Republican district and yields nine Democratic and six Republican seats under both sets of returns.\footnote{Andrew Green also submitted a plan with this seat split and a lower Democratic vote share in the seventh-most Republican district.}

Despite the range of outcomes among individual submissions, the median Democratic vote shares of the public plans and the 2025 ensemble are close at most district ranks. The clearest exception is the second-most Democratic district, where the public-plan median is near the upper end of the ensemble's distribution.

\section{Cycle Walk Navigates Ohio's Restrictive Plan Space}

Ohio presents a particularly challenging case for ensemble-based redistricting plan analysis because it has a complex system of rules on the division of political subdivisions. Some of these rules may be encoded as hard constraints and some as soft. We impose district contiguity and population balance as hard constraints, implementing rules of political subdivision as hard constraints severely restricts the free movement of Markov chains through the plan space. In our initial exploratory attempt, we ran four independent chains with all-hard constraints with a Multi-scale Metropolized Forest ReCom \cite{autry_metropolized_2021} whose initial design was motivated by preserving counties. In North Carolina, Multi-scale Metropolized Forest ReCom was fairly successful at sampling while preserving the counties \cite{mattingly_nc_expert_report_2021}. However even after extensive sampling, independently initialized chains show substantial persistent discrepancies across different chains in their marginal distributions in our Ohio setting. This is an indication of insufficient mixing. This experience motivated us to use the Cycle Walk algorithm, in which the system of political subdivision preservation is implemented by soft constraint and we used sub-sampling to enforce hard constraints.

Cycle Walk encourages the preservation of political subdivisions through weighted edges and energy penalties against ousted population. Edge weighting has also been used in ReCom, its Metropolized variants, and other methods to achieve the same goal \cite{duchin_spencer_models_2021,duchin_south_carolina_2022,tapp_parameter_2025,babson_models_2026,deford_cyclewalk_2025}. However, edge weighting alone is not enough: We have found that we must include additional energetic functions to match the non-partisan policies put forward by the Ohio redistricting commission. Cycle Walk is particularly accommodating to changes in the measure because it can mix large, small, and intermediate size moves when proposing new districts; the relative flexibility of additing additional measure information to Cycle Walk when compared with Metropolized Recombination. In \cite{deford_cyclewalk_2025}, adding certain types of energy to the measure are shown to quickly degrade mixing times in ReCom, whereas Cycle Walk is able to more flexibly adapt and efficiently sample a wider range of modified probability measures.

In the minimal compliant criteria ensemble we altered the edge weights to ensure that trees in the districting space do not meander in and out of Columbus, however, we also needed to add an ousted population energy to ensure those trees were cut at appropriate places. When using edge weights alone, we were not able to adequately control the amount of ousted population in Columbus. 

The Ohio Constitution grants mapmakers substantial discretion in the extent to which they preserve the municipal corporation and townships, with effort-based language like ``the authority shall attempt to include a significant portion of [a large] municipal corporation or township in a single district" \cite{oh_const_art_xix_sec2}. Flexibility in the probability distribution allows us to tune the number of split municipal corporations and townships and ousted population share to the comparable levels of the three official plans. We leverage this tool to obtain ensembles matching different official plans' varied municipality preservation policies.

Thanks to the soft constraints with subsampling, Cycle Walk chains more freely tour the highly restrained state space. As evidenced by the rank-normalized split-$\hat{R}$ statistics shown in Appendix~\ref{apdx:convergence}, within each ensemble, soft constraints empower Cycle Walk to attain marginal distributions of observations that highly agree across the independent chains. We have studied other measures of convergence such as pairwise total-variation distances and all provide strong evidence of convergence. The reliability in convergence of Cycle Walk across multiple chains justifies us pooling these samples in order to have a greater number of samples when performing electoral analysis.

\section{Discussion}

In our analysis, we have illustrated the considerable discretion available to the Ohio Redistricting Commission in interpreting and implementing the constitutional mandates and the nonpartisan redistricting criteria. Even among plans fully in compliance with the constitutional requirements, substantial variation exists in the extent to which political subdivisions are preserved. We constructed two ensembles tuned to match the political subdivision preservation policies of the 2021/22 and 2025 official plans, respectively. Through a third ensemble, we further showed that a better preservation of municipal corporations and townships is attainable.

Despite substantial differences in the geographic characteristics of the three ensembles, we find remarkably little variation in their electoral outcomes. The core lesson we learned in comparing the electoral consequences of the three ensembles is that increasing the degree of municipal preservation does not appear to systematically favor either political party. These observations are consistent with previous studies \cite{tapp_parameter_2025,wise_nosplit_2026}. One place in which our study differs from the others is that we have carefully matched the non-partisan criteria expressed by the Ohio Constitution and Redistricting Commission and have efficiently independently sampled representative sets of districts that show agreement across a range of relevant observables. Across all three ensembles, the official plans consistently stand out as outliers in many electoral observables. These observations unequivocally suggest that the partisan characteristics of the official plans cannot be attributed to the commission's policies of preservation of political subdivisions.

The commission, with flexibility in how it chooses to adopt certain criteria, had the opportunity to devise a plan that both preserves more counties and municipalities and more fairly responds to the will of the people. For example, compared with the official plans, as suggested by Figure~\ref{fig:cuts-and-ousted}, we find plans that displace a smaller share of population from municipal corporations and split fewer municipal corporations and townships. These same plans match similar levels of compactness and yield seat counts closer to proportionality and respond to vote share swing more symmetrically.
In Ohio, the objectives of preserving political subdivisions and limiting party favoritism do not appear to be in conflict.

The analysis of the November 2026 elections illustrates that the consequences of gerrymandering can sometimes extend beyond the outcomes of individual elections. Even when two redistricting plans produce the same number of Democratic seats, they may create very different conditions under which the two parties compete for representation. For example, with a statewide vote share in the elections in November 2026, as presented in Figure~\ref{fig:26-gov}, everyone in Ohio could expect to vote under districts that gradually change with the will of the people yet are now stuck in less responsive states on two sides of a firewall.

Beyond Ohio's congressional districts, the methodological framework developed in this work can be extended to Ohio's state legislature districts and other states' congressional districts. We are currently extending our analysis of Ohio to its State Senate and State House of Representatives districts, where the larger number of districts and the more complex system of constitutional requirements present additional challenges. We defer a comprehensive analysis of Ohio's General Assembly to a future paper.

There are a few limitations of this work. We present an evidence-based analysis of the three official plans of Ohio.
While we tune ensembles to the legislature's or the commission's policies of political subdivision preservation, it is an impossible task to reconstruct the decision-making process behind every decision in the official plans to split or preserve a political subdivision. There are are a host of other possible considerations when drawing policy based ensembles as well, such as whether or how much to allow double bunking of incumbant candidates.
We also do not claim to exactly capture the 2026 election data since it is still in the process of being published. However, we view historical elections as fixed and plausible voting patterns, rather than predictive, and thus can use them to to shed light on how partisan voting choices and voter turnout may affect electoral outcomes under changes to redistricting plans. We also use a base measure of weighted spanning trees which, although is consistent certain with exisiting policy goals, was likely not used as a direct policy consideration by the commission; specifically, compactness is a difficult concept to pin down in redistricting as it is stated without formal definition. There was some discussion and reporting on Polsby-Popper and Reock in the Ohio process \cite{cooper2022compactness,league2022opinion,rodden2023comparison,freilich2021testimony,ohio2021transcript}, however, it isn't entirely clear if this was what the commission actually had in mind either. Given the consistency of our ensembles with the commissions plans in terms of these measure, we feel we have done adequate due-diligence, however, it is still an open question as to (i) how to build evidence for policy based on observed maps and (ii) how to continue to expand the class of measures that may be efficiently sampled. There is recent work exploring the latter problem, which includes the use of the Cycle Walk algorithm (e.g. see \cite{deford_cyclewalk_2025,chuang2025multiscale,wang2026stratifiedsamplingredistrictingplans}).

\section{Conclusion}

This work presents a comprehensive analysis of the three official congressional redistricting plans in Ohio. It also demonstrates that Cycle Walk is a powerful yet flexible algorithm for generating redistricting ensembles under a highly complex system of geographic constraints; originally Cycle Walk was tested on a variety of cases, but never with the full set of criteria of an actual redistricting commission.
We show that all three official plans, while structurally comparable to the ensemble, are extreme outliers in terms of electoral outcomes.
As a deviation from proportionality and symmetry, the three official plans consistently produce Republican-favored outcomes based on the historical vote share.

The case of Ohio is a strong testament to the power of Cycle Walk in generating redistricting ensembles in a flexible manner.
The state's political subdivisions, in practice, cannot be simultaneously preserved; even the official plans have to make trade-offs between preserving counties and municipalities.
The Cycle Walk algorithm is our translation of these necessary trade-offs into a mathematical framework.
Through weighted edges, Cycle Walk strongly encourages the preservation of counties and municipalities, while still allowing for the exploration of the space of redistricting plans.
The weights of county and municipal edges can even be adjusted for us to engineer a particular regime that leads to a set of structural and municipal preservation metrics.
This flexibility allows us to generate ensembles that overlap with the official plans in these metrics and thus allow for a fair comparison of electoral outcomes.

\section{Acknowledgments}

During the preparation of this work, the authors were assisted by several Generative AI tools, especially ChatGPT 6 Astra, ChatGPT 6.2 Sol (OpenAI), and Claude Opus 5.5 (Anthropic) for many details of computational work. The paper was written by us with copyediting assistance from AI. We assume full responsibility for all mathematical claims presented, and all results have been independently verified by the authors.

\appendix

\section{Other Run Parameters}
\label{apdx:runparams}

We report additional runtime parameters used in the ensemble generation. When using Cycle Walk, we perform 9 1-tree cycle walks per every 2-tree cycle walk.

For weighting cross-county edges, we initially set the cross county edge weights in the graph to 1/50; due to a quirk of our run scripts, we scaled these weights by the city/village edge weight, $w_{CV}$, meaning that the cross county weights ended up being 0.1, 0.2, 0.2, 0.4 for the four ensembles with $w_{CV}$ set to be 5, 10, 10, and 20, for the minimal criteria, 2025 criteria, 2021/2022 criteria, and high preservation criteria, respectively.

The purpose of the cross county weights is to achieve hard constraints via sub-sampling, and we found that each ensemble was able to find an adequate number of samples that adhered to the hard county constraints. We find that de-emphasizing cross county weights is essential for county preservation, however, there appears to be an insensitivity to precisely how much we de-emphasize these connections.

We collect our ensembles on the Duke University Department of Mathematics Sun Grid Engine compute cluster.
The computational tasks are randomly assigned to one of the 16 cluster machines; all cluster machines currently have Quad Core Intel i5 CPUs running at 3.3 GHz, 8GB of RAM. We check a plan for compliance every 1000 steps of the Markov Chain (which is roughly 100 2-tree Cycle Walk steps). The minimal criteria chain samples at over 25,000 steps per second (or over 2,500 2-tree Cycle Walk steps per second). When we check for compliance to hard constraints, the  minimal criteria ensemble has a 1.326\% compliant rate.
The 2021/2022 criteria was sampled at a rate of 10,978 steps per second with a 9.07\% compliance rate.
The 2025 criteria was sampled at a rate of 10,320 steps per second with a 3.58\% compliance rate.
The high preservation criteria was sampled at a rate of 9,924 steps per second with a 0.862\% compliance rate.

\section{Sampling Convergence}
\label{apdx:convergence}

We ran four independent Cycle Walk chains for each ensemble and obtained 20,000 plans (via subsampling plans that adhere to the hard county constraints).
Before pooling the four chains within each sampling configuration, we evaluated whether those chains produced consistent distributions of selected observables.

To test for sampling convergence we use order statistics on the compactness of each district. To specify, we examine the isoperimetric ratio (which is proportional to the inverse of the Polsby-Popper score) for each of the 15 districts in a given plan and then sort these ratios from least to greatest. Repeating this across plans, gives a marginal density of the most compact district, second most compact district, etc.

We repeat this analysis by using order statistics on partisan lean of each district according to 2020 US Presidential statewide vote counts. The partisan lean is given by the Democratic votes in a district divided by the Democratic and Republican votes (the numerator is interchangeable with the Republican votes), and we then sort the 15 districts in a plan from the highest to lowest fraction. We then get marginal distributions for the most Democratic district, the second-most Democratic district, and so on through the most Republican district.

For each of the 30 ordered marginal densities, we examine the Gelman-Rubin statistic for the expected value of the density, using the rank-normalized split-$\hat{R}$ method \cite{gelman_inference_1992, vehtari_rank_2021}.
Unlike the classic Gelman-Rubin statistic, this method pools all chains, ranks the samples from the largest to the smallest, partitions all chains into two halves, and replaces each raw value with its pooled rank. Splitting chains into two halves enhances the sensitivity of non-stationarity among chains, while rank-normalization process mitigates the influence of extreme observations on the estimate of variance. These improvements are particularly useful in our redistricting ensemble because many (district-level) observables usually exhibit skewed distributions, heavy tails, or slow mixing.
We report the average $\hat{R}$ and max $\hat{R}$ as in Table~\ref{tab:rhat-diagnostics}. Thus all four ensembles exhibit good convergence properties.

\begin{table}[htbp]
    \centering
    \caption{Rank-normalized split-$\hat{R}$ statistics across four ensembles.}
    \label{tab:rhat-diagnostics}
    \begin{tabular}{lcc}
        \toprule
        Ensemble & Average $\hat{R}$ & Max $\hat{R}$ \\
        \midrule
        Minimal Criteria       & 1.0004 & 1.0011 \\
        Tuned to 2025          & 1.0025 & 1.0068 \\
        Tuned to 2021/22       & 1.0057 & 1.0189 \\
        High Preservation      & 1.0029 & 1.0071 \\
        \bottomrule
    \end{tabular}
\end{table}

In addition to evaluating the Gelman-Rubin test, we also examine the pairwise total-variation distance between similarly ranked order marginals across independent chains.
We estimate the total variation by creating histograms. The bin width for the isoperimetric ratio is 0.5 and for the partisan lean is 0.2\%. We keep aligned bins to have a consistent comparison across all ensembles.
The total-variation distance is computed this way for each pair of chains between the distributions at each district rank and averaged these distances across the 15 ranks. We report the largest average among the six pairs of chains and examine how this quantity changes with the number of saved plans.\\

\begin{figure}[htbp]
    \centering

    \begin{subfigure}[b]{0.49\textwidth}
        \centering
        \includegraphics[width=\linewidth]{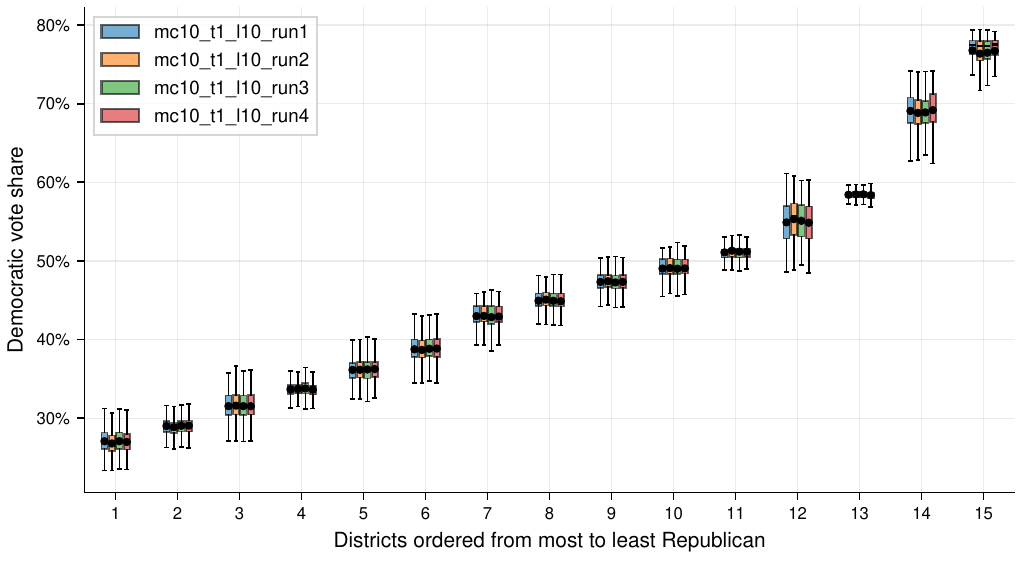}
        \caption{Ordered Democratic vote-share marginal distribution box plot.}
        \label{fig:match25-vote-marginals}
    \end{subfigure}
    \hfill
    \begin{subfigure}[b]{0.49\textwidth}
        \centering
        \includegraphics[width=\linewidth]{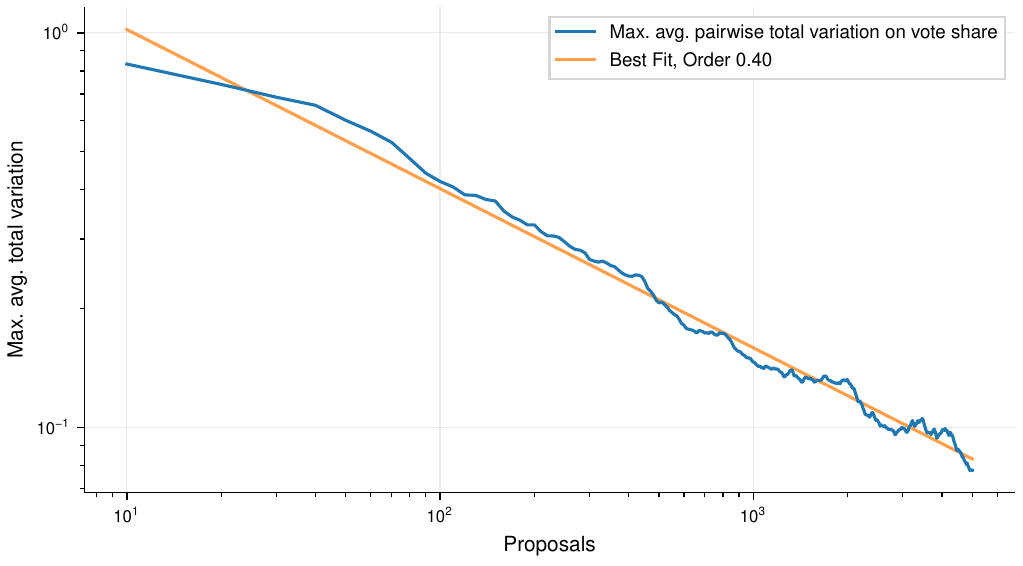}
        \caption{Pairwise total-variation convergence for Democratic vote share.}
        \label{fig:match25-vote-tv}
    \end{subfigure}

    \vspace{0.5cm}

    \begin{subfigure}[b]{0.49\textwidth}
        \centering
        \includegraphics[width=\linewidth]{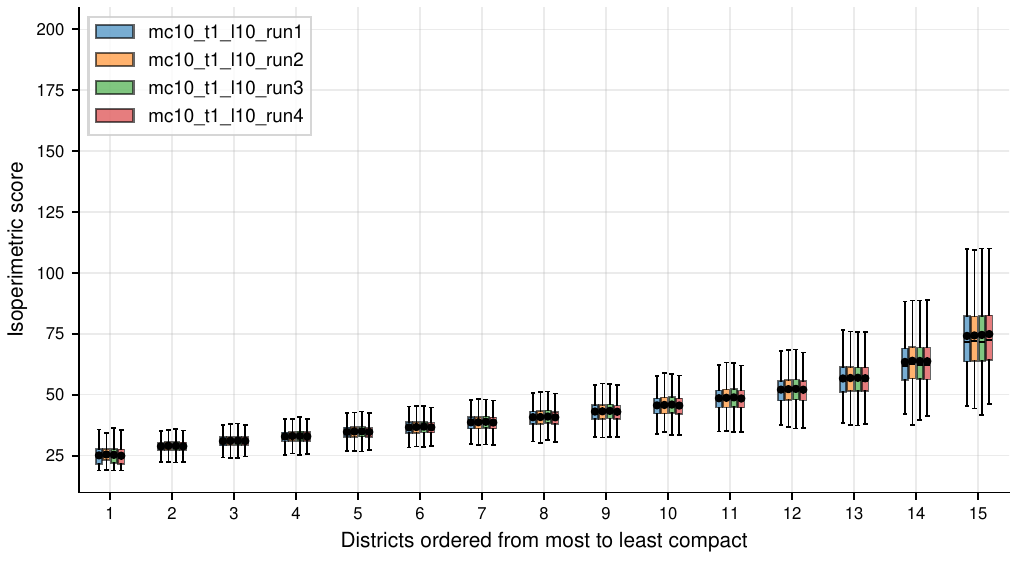}
        \caption{Ordered isoperimetric-ratio marginal distribution box plot.}
        \label{fig:match25-compactness-marginals}
    \end{subfigure}
    \hfill
    \begin{subfigure}[b]{0.49\textwidth}
        \centering
        \includegraphics[width=\linewidth]{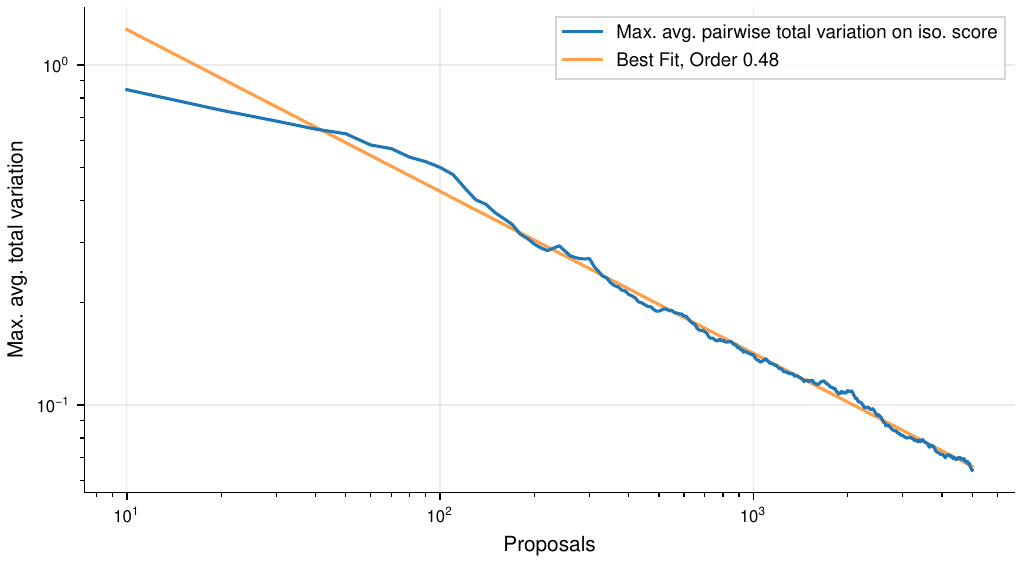}
        \caption{Pairwise total-variation convergence for isoperimetric ratio.}
        \label{fig:match25-compactness-tv}
    \end{subfigure}

    \caption{
    Convergence diagnostics for the 20,000-plan ensemble matching municipality preservation of Enacted 2025 Plan.
    Panels (a) and (c) compare the ordered district-level marginal
    distributions after retaining 5,000 compliant plans; the maximum average pairwise
    total-variation distances are 0.0779 for Democratic vote share and
    0.0644 for isoperimetric ratio, respectively.
    Panels (b) and (d) show the corresponding decrease in total-variation
    distance over the course of the runs, with fitted power-law exponents
    of approximately 0.40 and 0.48, respectively.
    }
    \label{fig:match25-convergence}
\end{figure}

We demonstrate differences in quantiles of the order statistics across the four independent chains for the ensemble with the 2025 matched criteria in \Cref{fig:match25-convergence}. We also show how the worst case total variation, as averaged across the 15 statistics, reduces as a function of the number of samples.

\section{Estimation of Precinct-level Two Party Votes in November 3, 2026 Elections}
\label{apdx:2026}

Shortly after the November 3, 2026 elections, we compiled Democratic and Republican vote counts for eight statewide contests. The collection concluded on November 21, 2026. We prioritized certified precinct-level official canvasses, then unofficial precinct-level canvasses, and used countywide returns where precinct-level results were not available. Among Ohio's 88 counties, we collected official precinct returns from 55 counties, unofficial precinct returns from 20 counties, official countywide totals from 4 counties (Jackson, Muskingum, Sandusky, and Trumbull), and unofficial countywide totals from 9 counties (Columbiana, Gallia, Hancock, Highland, Lucas, Mahoning, Medina, Pike, and Williams).

Since our sampling process is based on the 2020 precinct graph produced by VEST \cite{vest_2020}, we matched 2026 reporting precincts to the 2020 precincts using names, county boundary files, and public records of precinct changes. Where a reporting precinct mapped to a single graph node, we assigned its vote count intact. When a 2026 reporting precinct contained territory from more than one 2020 precinct, we estimated what share of each 2020 precinct's voting-age population lived within the 2026 boundary. We multiplied each party's 2020 presidential vote count in each precinct by that share. We then distributed each party's reported 2026 votes among the 2020 precincts in proportion to those adjusted counts. When we had a 2026 precinct's vote totals and could identify several possible matched 2020 precincts, but did not have sufficient information to measure the overlap, we split the votes among the group of 2020 precincts in proportion to their respective votes in 2020 Presidential Election. The same method applies to the case where a county's precinct-level report is not available. Whenever we have to allocate the votes by fractions, we floor each allocation and assign the remaining votes to nodes with the largest fractional remainder.

The original data of our estimation of precinct-level two party votes in the eight statewide races in 2026 elections can be found at \texttt{https://github.com/franklinxzhou/Ohio-Election-Data-2026.git}. We note that although the data used to represent the 2026 elections is not exact, we use all of the voting patterns to provide plausible and fixed voting scinarios so that we may understand the impact of redistricting while keeping the votes fixed.

\section{Electoral Analysis of Other Ensembles}
\label{apdx:electoral-analysis-other}

In this appendix, we present the corresponding electoral analyses for the Match 2021/22 and High Preservation ensembles.
These results complement the electoral analysis of the Match 2025 ensemble presented in the main text (Figure~\ref{fig:ceah-zoom} and Figure~\ref{fig:asymmetry-histogram}) and illustrate the extent to which our electoral findings persist across different policies of political subdivision preservation.

\begin{figure}[htbp]
    \centering
    \includegraphics[width=1\textwidth]{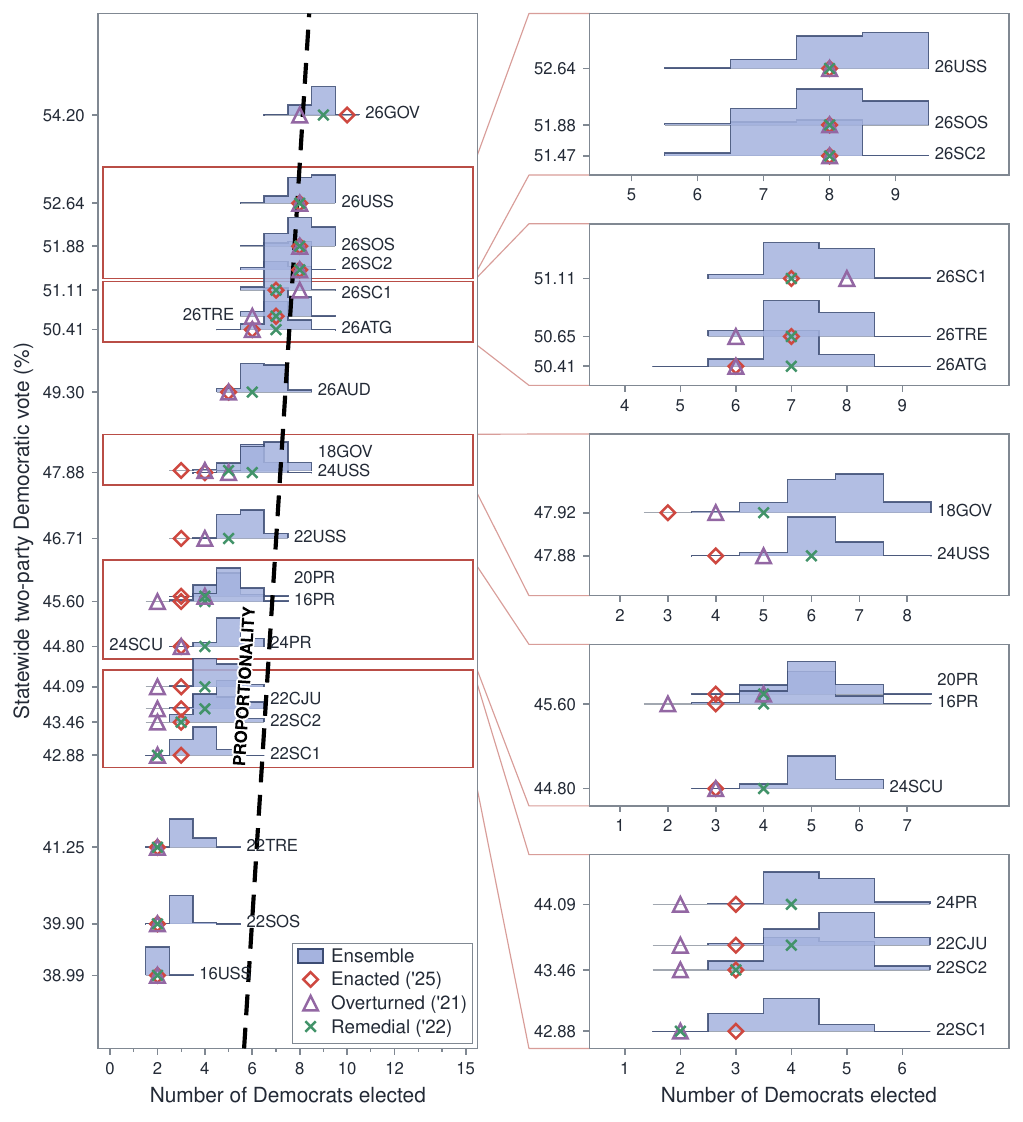}
    \caption{Collected Seat Histogram of the number of Democratic seats in the high preservation ensemble, based on selected historical statewide elections in Table~\ref{tab:elections}. In the Y-axis, the historical elections are ordered according to the Democratic vote share. The colored dots on the X-axis represent the number of Democratic seats based on official plans. The dashed lines represent the Democratic seat counts based on strict proportional representation. Election labels, see Figure~\ref{fig:ceah-zoom}.}
    \label{fig:ceah-zoom-hp}
\end{figure}

\begin{figure}[htbp]
    \centering
    \includegraphics[width=0.45\textwidth]{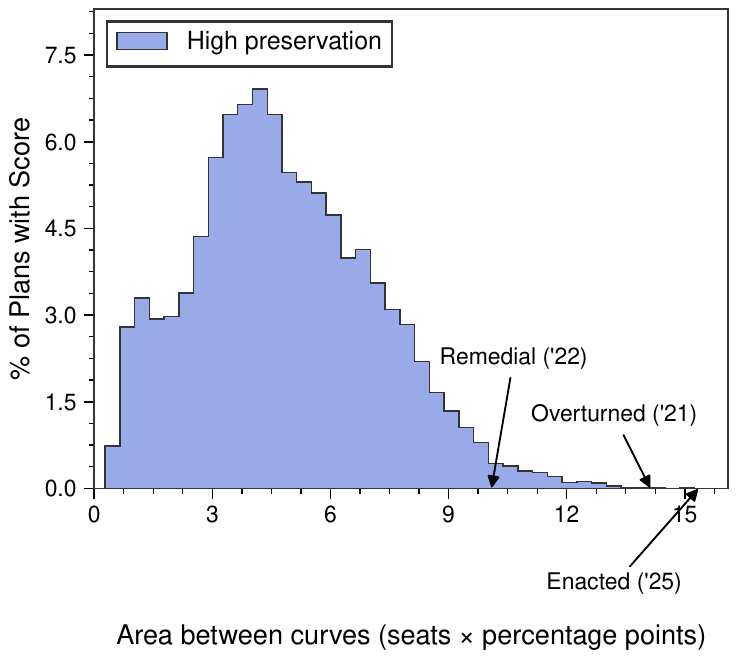}
    \caption{Distribution of partisan asymmetry scores among 20,000 sampled plans in the high preservation ensemble. The score is defined as the area between the reciprocal seats-votes curves, averaged across elections in Table~\ref{tab:elections}, over the range in which the respective party receives 50-55\% of the statewide two-party vote.}
    \label{fig:asymmetry-hp}
\end{figure}

\begin{figure}[htbp]
    \centering
    \includegraphics[width=1\textwidth]{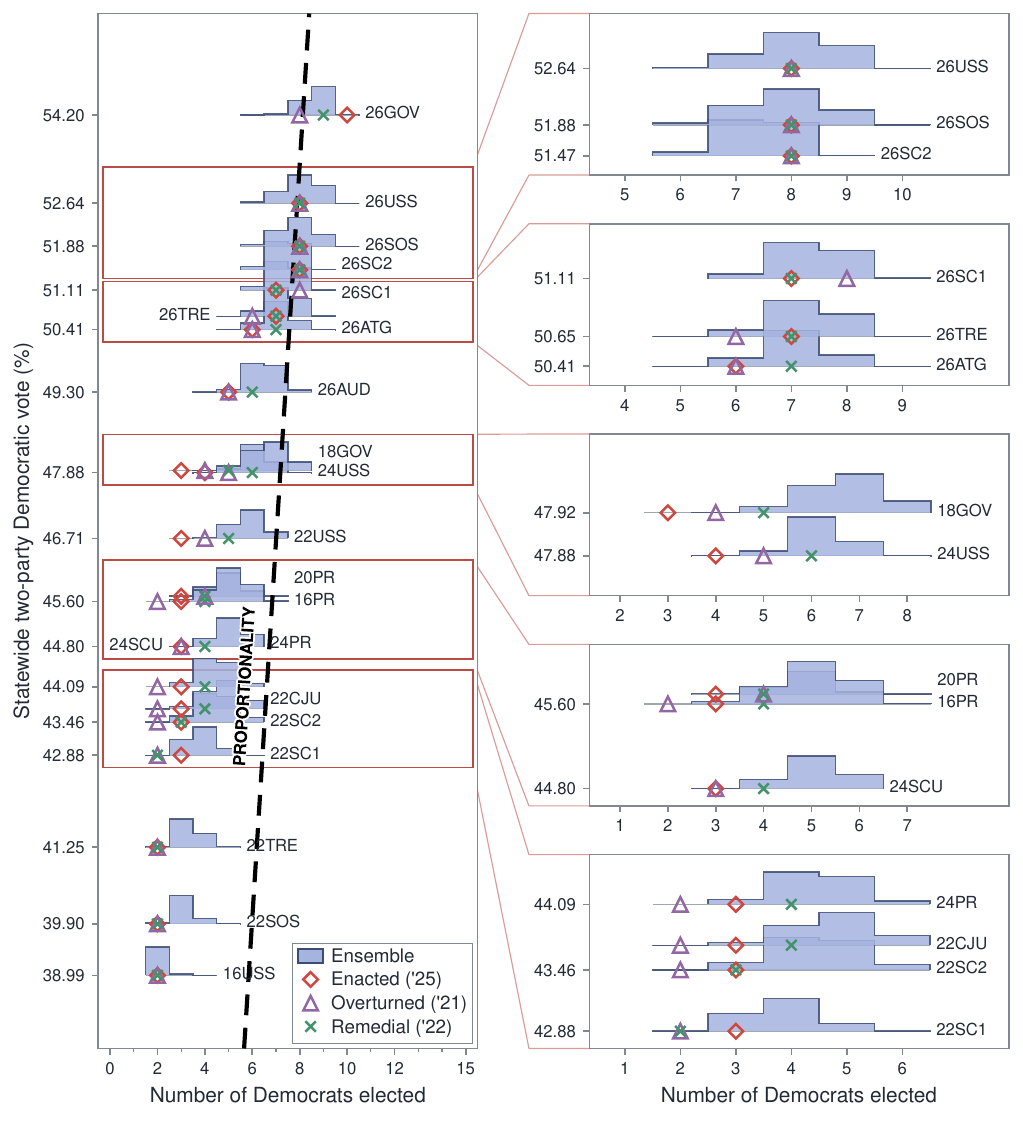}
    \caption{Collected Seat Histogram of the number of Democratic seats in the ensemble tuned to match the policy of municipality preservation of 2021 Overturned/2022 Remedial plans, based on selected historical statewide elections in Table~\ref{tab:elections}. In the Y-axis, the historical elections are ordered according to the Democratic vote share. The colored dots on the X-axis represent the number of Democratic seats based on official plans. The dashed lines represent the Democratic seat counts based on strict proportional representation. Election labels, see Figure~\ref{fig:ceah-zoom}.}
    \label{fig:ceah-zoom-match2122}
\end{figure}

\begin{figure}[htbp]
    \centering
    \includegraphics[width=0.45\textwidth]{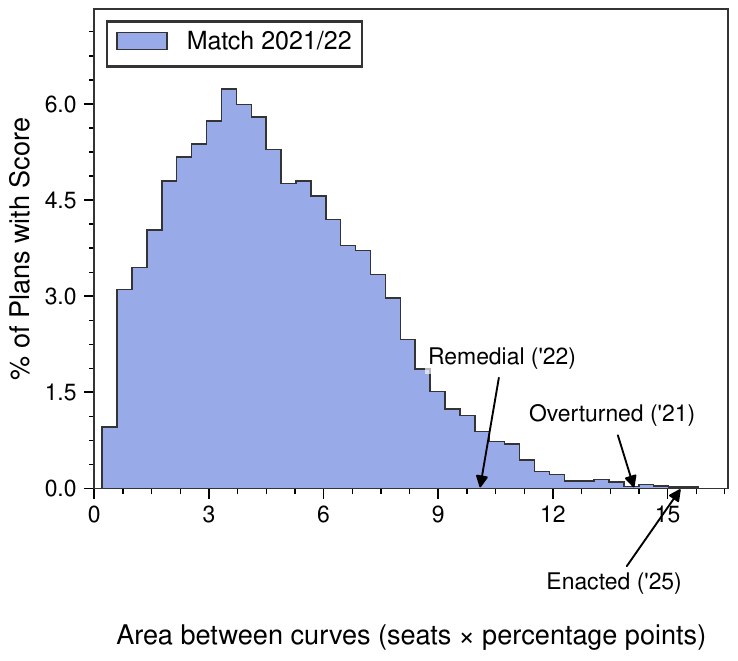}
    \caption{Distribution of partisan asymmetry scores among 20,000 sampled plans in the ensemble tuned to match the policy of municipality preservation of 2021 Overturned/2022 Remedial plans. The score is defined as the area between the reciprocal seats-votes curves, averaged across elections in Table~\ref{tab:elections}, over the range in which the respective party receives 50-55\% of the statewide two-party vote.}
    \label{fig:asymmetry-match-2021-22}
\end{figure}

\bibliographystyle{plain}
\bibliography{references}

\end{document}